\documentclass[aps,prd,preprint,superscriptaddress]{revtex4}

\usepackage{graphicx}
\usepackage{amsmath}
\usepackage{bm}
\usepackage{color}
\usepackage{amssymb,enumerate}

\def\ee{\end{eqnarray}}

\def\p{\partial}

\def\=:{=\hspace{-.7em}\raisebox{1.1ex}{.}\hspace{.1em}\raisebox{-0.2ex}{.} }

\def\ee{\end{eqnarray}}

\def\p{\partial}

\def\=:{=\hspace{-.7em}\raisebox{1.1ex}{.}\hspace{.1em}\raisebox{-0.2ex}{.} }

\newcommand {\beq}{\begin{eqnarray}}
\newcommand {\eeq}{\end{eqnarray}}
\newcommand {\non}{\nonumber\\}

\newcommand {\1}[1]{\frac{1}{#1}}

\newcommand {\ph}{\varphi}

\newcommand {\del}{\partial}

\newcommand {\tr}{{\rm tr}\,}

\begin{document}


\title{
Josephson instantons and Josephson monopoles\\
in a non-Abelian Josephson junction
}


\author{Muneto Nitta}

\affiliation{
Department of Physics, and Research and Education Center for Natural 
Sciences, Keio University, Hiyoshi 4-1-1, Yokohama, Kanagawa 223-8521, Japan\\
}


\date{\today}
\begin{abstract}
Non-Abelian Josephson junction is 
a junction of non-Abelian color superconductors sandwiching an insulator, 
or a non-Abelian domain wall if flexible,   
whose low-energy dynamics is described by
a $U(N)$ principal chiral model with the conventional pion mass.
A non-Abelian Josephson vortex is 
a non-Abelian vortex (color magnetic flux tube) 
residing inside the junction, that is described as  
a non-Abelian sine-Gordon soliton.  
In this paper, we propose Josephson instantons and Josephson monopoles,　that is, Yang-Mills instantons and monopoles inside a non-Abelian Josephson junction, respectively, and show that they are described as  
$SU(N)$ Skyrmions and $U(1)^{N-1}$ vortices 
in the $U(N)$ principal chiral model without and with a twisted mass term, respectively.  
Instantons with a twisted boundary condition 
are reduced  (or T-dual) to monopoles,  implying that 
${\mathbb C}P^{N-1}$ lumps are T-dual to 
${\mathbb C}P^{N-1}$ kinks inside a vortex. Here we find 
$SU(N)$ Skyrmions are T-dual to $U(1)^{N-1}$ vortices inside a wall. 
Our configurations suggest a yet another  
duality between  ${\mathbb C}P^{N-1}$ lumps 
and $SU(N)$ Skyrmions as well as that between ${\mathbb C}P^{N-1}$ kinks and $U(1)^{N-1}$ vortices, 
viewed from different host solitons. 
They also suggest a duality between fractional instantons and bions in the ${\mathbb C}P^{N-1}$ model and 
those in the $SU(N)$ principal chiral model.

\end{abstract}

\maketitle

\section{Introduction}

Yang-Mills instantons and magnetic monopoles 
are two  topological solitons 
studied very well in both physics and mathematics 
\cite{Manton:2004tk}:
they are both integrable, 
admit hyper-K\"ahler moduli spaces, 
and their solutions are available through 
by the Atiyah-Drinfeld-Hitchin-Mannin \cite{Atiyah:1978ri} 
and Nahm \cite{Nahm:1979yw} constructions, respectively.  
They are related by a so-called Nahm transformation 
that can be now understood as a T-duality 
in D-brane realization of these objects in type-II string theory 
\cite{Giveon:1998sr}. 
A new twist on these objects found recently 
was their realizations in the Higgs phase 
in which gauge symmetry is completely broken 
when gauge fields are coupled with 
several Higgs scalar fields in 
the fundamental representation. 
In the Higgs phase, 
there can exist
a non-Abelian vortex
that has ${\mathbb C}P^{N-1}$ moduli 
 \cite{Hanany:2003hp,Auzzi:2003fs,Eto:2005yh} 
and a non-Abelian domain wall carrying 
$U(N)$ moduli \cite{Shifman:2003uh,Eto:2005cc,Eto:2008dm}. 
Although instantons cannot exist stably 
in the Higgs phase,
$SU(N)$ instantons can stably exist as 
${\mathbb C}P^{N-1}$ lumps (or instantons) \cite{Polyakov:1975yp} 
inside a non-Abelian vortex \cite{Eto:2004rz,Fujimori:2008ee}  
and as $SU(N)$ Skyrmions
inside a non-Abelian domain wall  \cite{Eto:2005cc}. 
(The latter setting physically realizes the 
Atiyah-Manton construction of Skyrmions 
from instanton holonomy \cite{Atiyah:1989dq}.)
On the other hand,
monopoles are confined by magnetic fluxes 
in the Higgs phase, 
and they become ${\mathbb C}P^{N-1}$ kinks 
\cite{Abraham:1992vb,Gauntlett:2000ib,Isozumi:2004jc,Isozumi:2004vg} 
inside a non-Abelian vortex 
\cite{Tong:2003pz,Shifman:2004dr,Hanany:2004ea,Nitta:2010nd}. 
Instantons with a twisted boundary condition
are reduced  (or T-dual) to monopoles, 
that is known as 
the Scherck-Schwartz (or twisted) 
dimensional reduction \cite{Scherk:1979zr}.
This implies inside a vortex in the Higgs phase that  
${\mathbb C}P^{N-1}$ lumps (instantons) 
with a twisted boundary condition 
are reduced (T-dual) to 
${\mathbb C}P^{N-1}$ kinks \cite{Eto:2004rz}. 
See Refs.~\cite{Tong:2005un,Eto:2006pg,Eto:2005sw,
Shifman:2007ce} 
as a review of these composite topological solitons. 
It is, however, not known thus far 
what it becomes if a monopole 
resides inside a non-Abelian domain wall.

We further pursue 
relations of among these topological solitons 
to find a complete circle.  
A key ingredient is a recently proposed 
non-Abelian Josephson junction 
\cite{Nitta:2015mma}, that is
a junction of 
non-Abelian color superconductors 
sandwiching an insulator,
or a non-Abelian domain wall if it is flexible. 
As for color superconductors, one can consider
either those in dense quark matter at high baryon density
\cite{Alford:2007xm,Eto:2013hoa}, 
or those in supersymmetric gauge theories in the Higgs phase 
\cite{Tong:2005un,Eto:2006pg,Eto:2005sw,
Shifman:2007ce}.
The low-energy dynamics 
of the non-Abelian Josephson junction 
can be described by the $U(N)$ principal chiral model \cite{Eto:2005cc}, 
in which the Josephson term in the bulk 
induces a pion mass term \cite{Nitta:2015mma}. 
When a non-Abelian vortex 
(or color magnetic flux tube)  
exists in the bulk color superconductor, 
it is absorbed into the junction if it exists.
The non-Abelian vortex 
residing inside the junction is referred as
a non-Abelian Josephson vortex (or fluxon)
 \cite{Nitta:2015mma}, 
that can be described as  
a non-Abelian sine-Gordon soliton 
\cite{Nitta:2014rxa} 
in the $U(N)$ principal chiral model
with the pion mass term.  
This is a non-Abelian extension of 
a Josephson vortex 
described by the usual sine-Gordon soliton 
\cite{Nitta:2012xq,Auzzi:2006ju}
in a Josephson junction of 
metallic superconductors \cite{Ustinov:1998}.
This correspondence was generalized to 
higher dimensional Skyrmions 
\cite{Nitta:2012wi} 
and to Yang-Mills instantons 
\cite{Nitta:2013cn,Nitta:2013vaa}.

\begin{figure}
\begin{center}
\begin{tabular}{ccccc}
\includegraphics[width=0.17\linewidth,keepaspectratio]{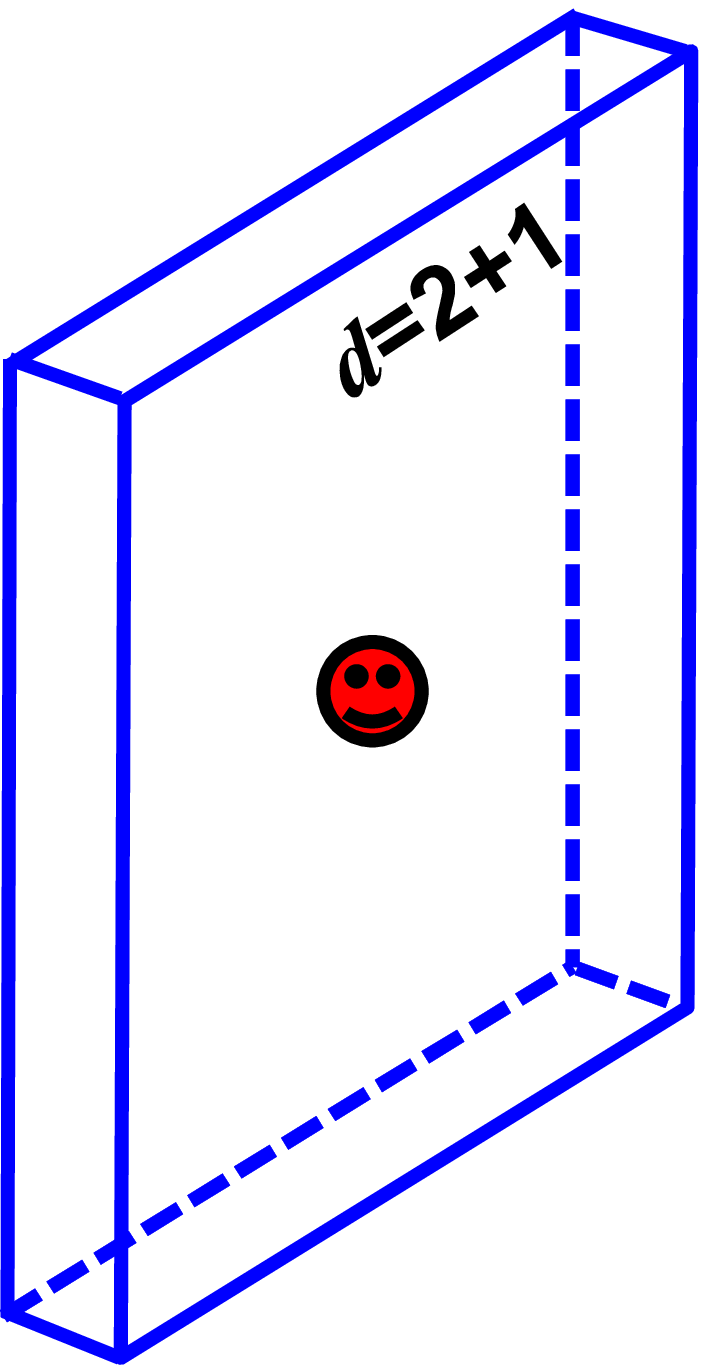}
& 
\includegraphics[width=0.45\linewidth,keepaspectratio]{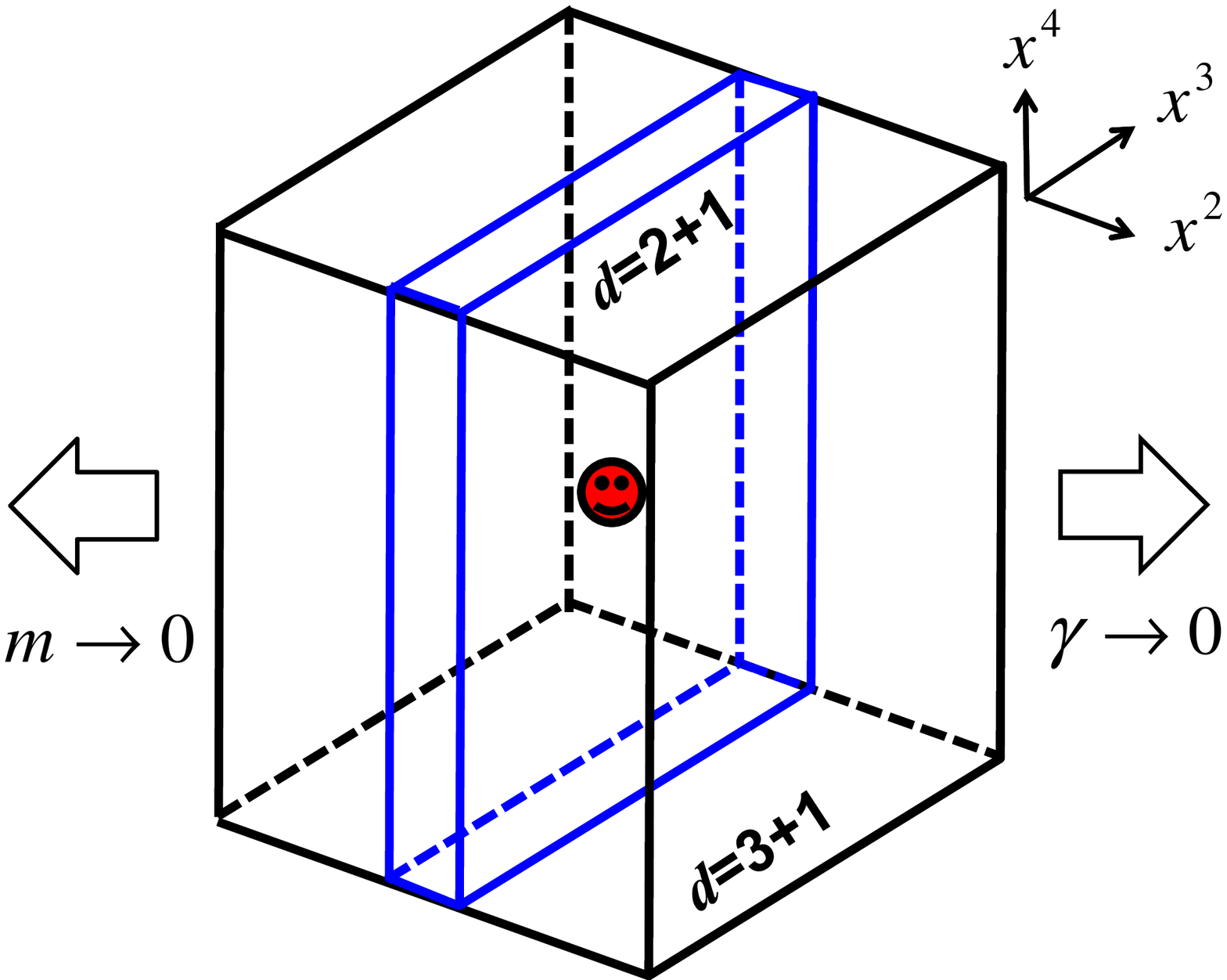} 
&
\includegraphics[width=0.28\linewidth,keepaspectratio]{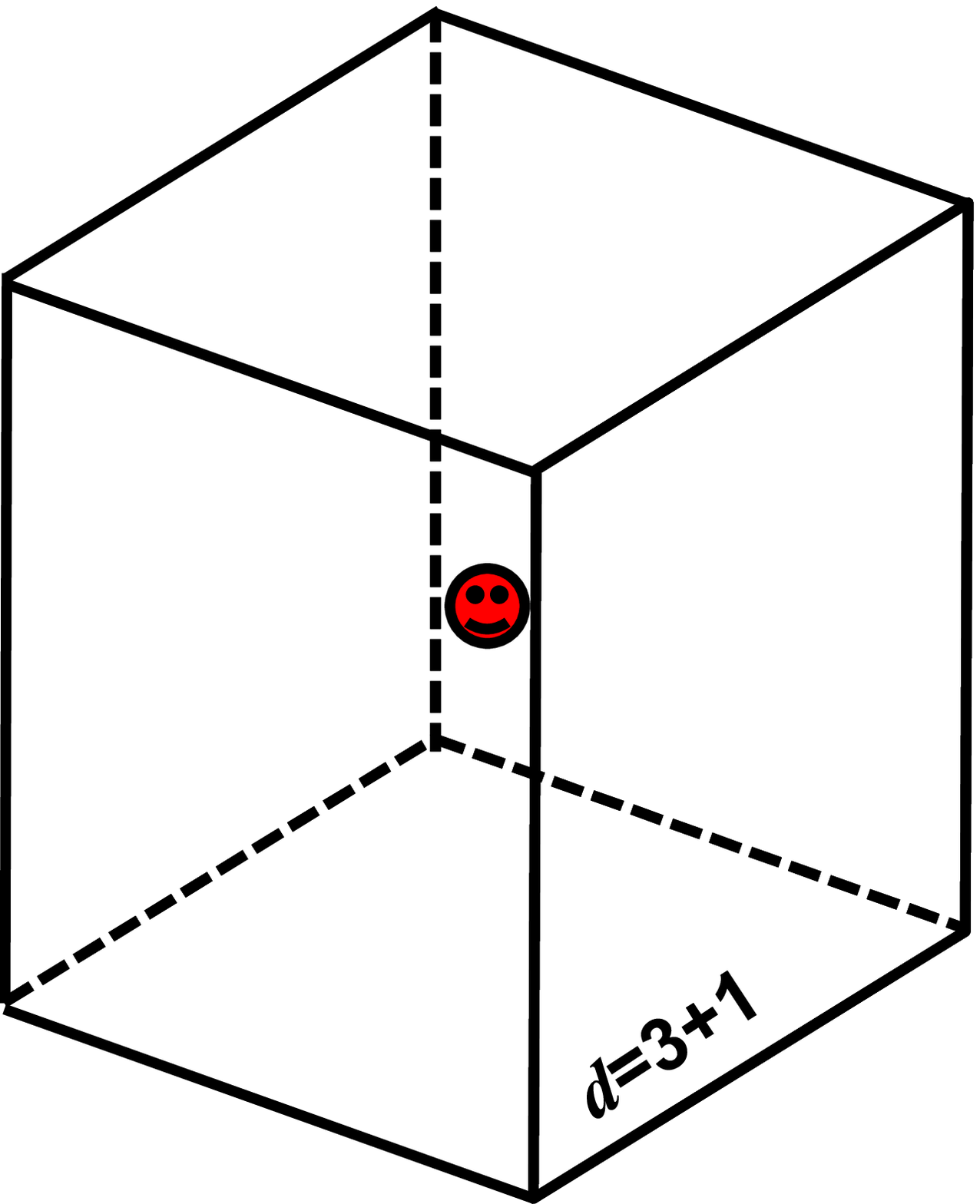}
\\
{\bf vortex-instanton} & {\bf wall-vortex-instanton} & {\bf wall-instanton}\\
(a) &  (b) &  (c) \\
\includegraphics[width=0.08\linewidth,keepaspectratio]{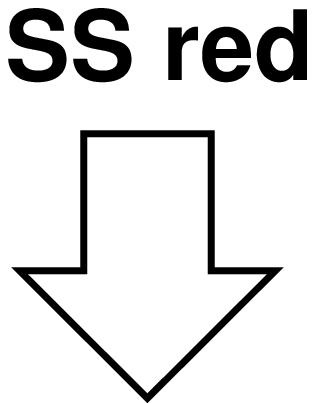} & 
\includegraphics[width=0.08\linewidth,keepaspectratio]{SS-red} & 
\includegraphics[width=0.08\linewidth,keepaspectratio]{SS-red} \\
\includegraphics[width=0.17\linewidth,keepaspectratio]{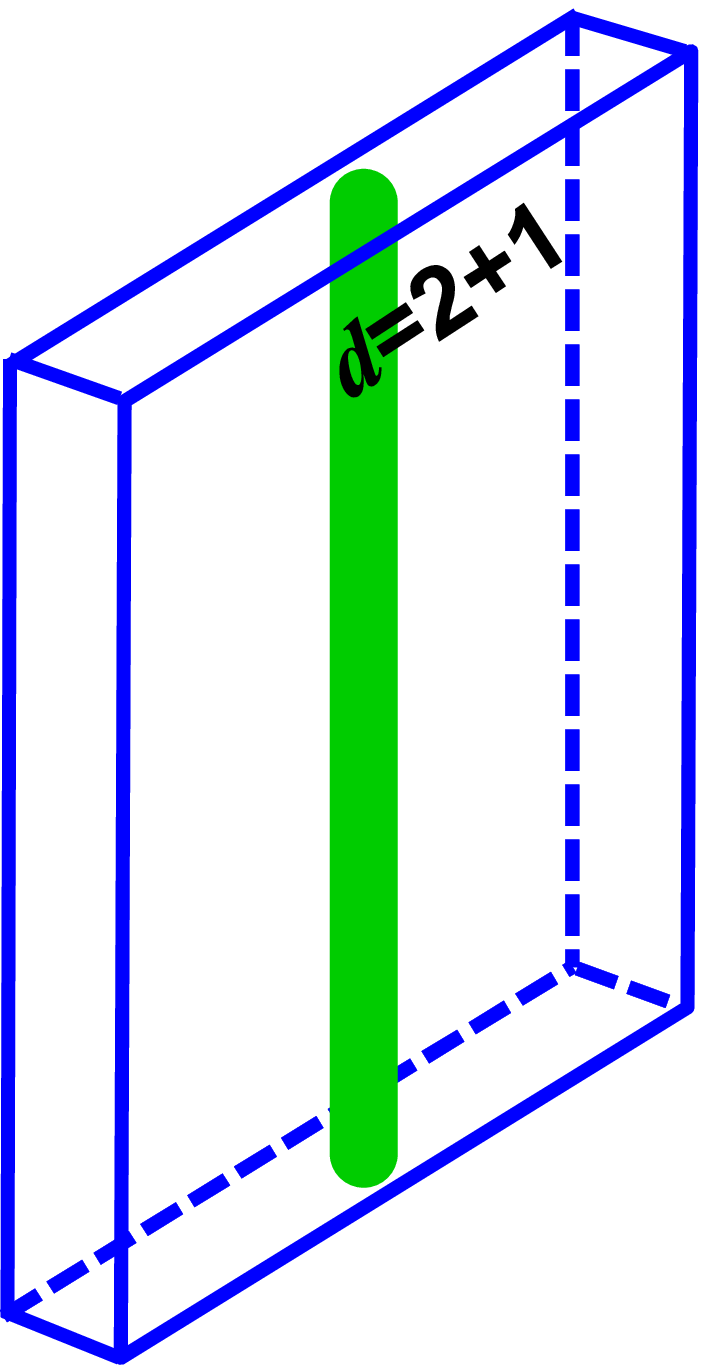}
&
\includegraphics[width=0.45\linewidth,keepaspectratio]{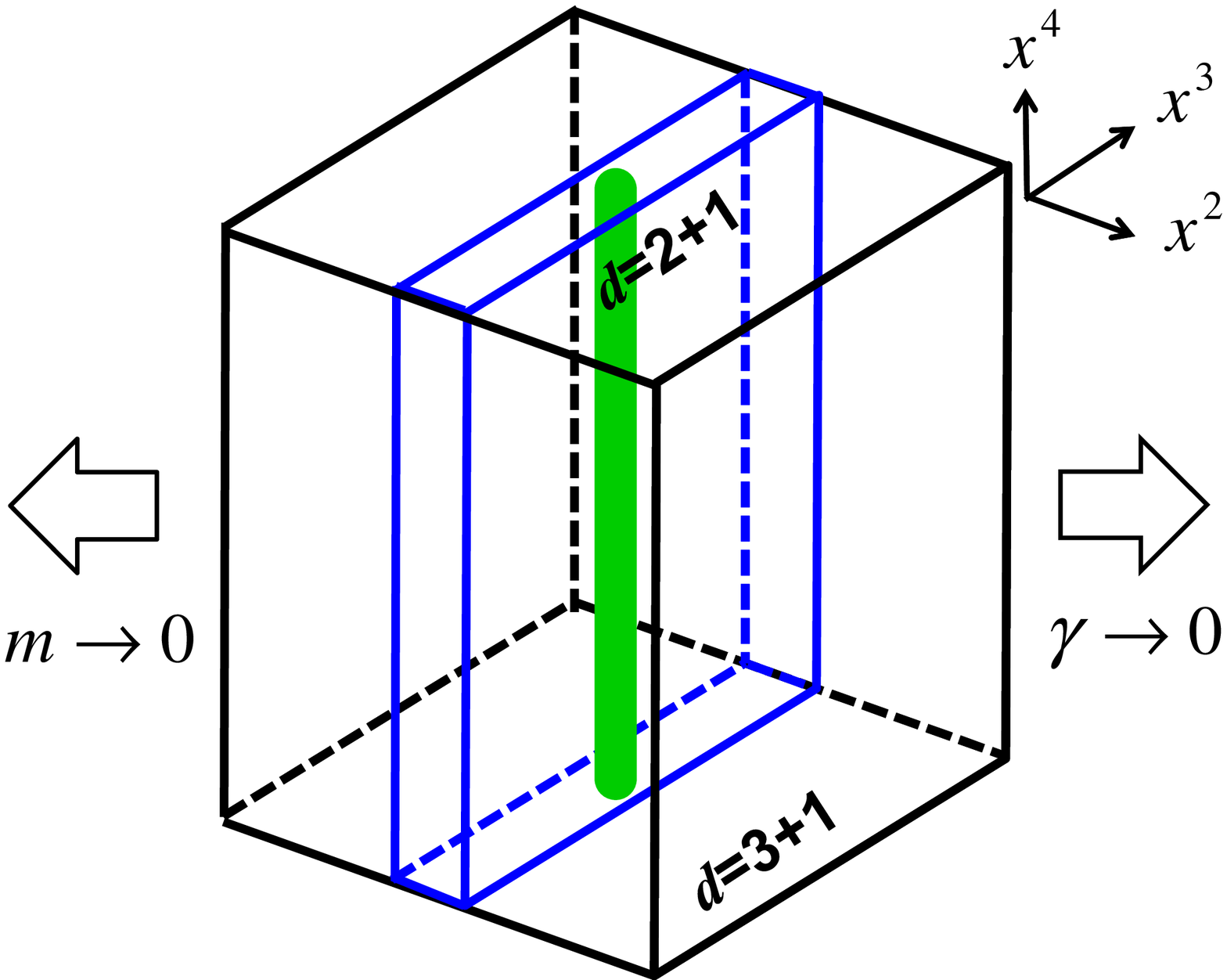} 
&
\includegraphics[width=0.28\linewidth,keepaspectratio]{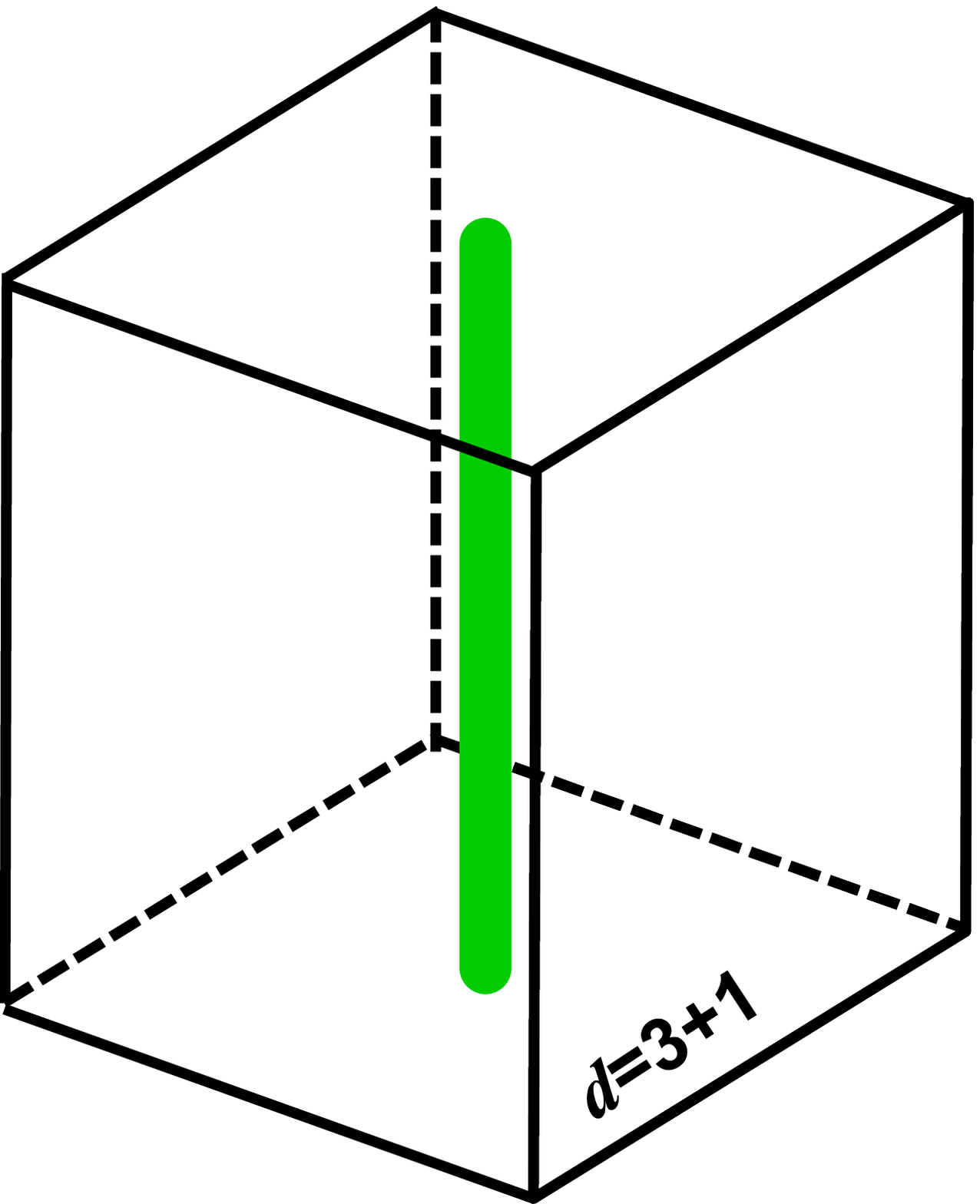} 
\\
{\bf vortex-monopole} & {\bf wall-vortex-monopole} & {\bf wall-monopole}\\
(d) &  (e) &  (f) 
\end{tabular}
\caption{Duality relations among all configurations.
The codimensional direction $x^1$ 
of the non-Abelian domain wall is not shown here.
The  (black) square boxes denote non-Abelian domain walls(Josephson junctions), 
(blue) thin boxes are vortices, 
(red) circles are 
Yang-Mills instantons, 
and (green) lines denote monopoles.
(a) Instanton inside a non-Abelian vortex, 
(b)  Instanton inside a non-Abelian vortex trapped in 
non-Abelian domain wall  (c) Instanton inside a non-Abelian domain wall (Josephson junction), 
(d) Monopole inside a non-Abelian vortex, 
(e) Monopole inside a non-Abelian vortex trapped in 
non-Abelian domain wall  
(f) Monopole inside a non-Abelian domain wall (Josephson junction). 
(c), (d) and (e) are obtained from 
(a), (b) and (c),  respectively, 
by the Scherk-Schwarz dimensional 
reduction or a T-duality.
}
\label{fig:monopole-instanton}
\end{center}
\end{figure}

In this paper, we propose 
Josephson instantons and Josephson monopoles,
that is, Yang-Mills instantons and monopoles inside a non-Abelian 
Josephson junction, respectively, and clarify their relations. 
We first construct Josephson instantons and monopoles 
residing a non-Abelian Josephson vortex inside 
the non-Abelian Josephson junction.
If  we remove the junction by taking massless limit of the Higgs fields, 
the configurations go to instantons
and monopoles inside the non-Abelian vortex, 
that is, instanton-vortex 
\cite{Eto:2004rz,Fujimori:2008ee} 
and monopole-vortex 
\cite{Tong:2003pz,Shifman:2004dr,Hanany:2004ea,Nitta:2010nd} 
composites known before. 
Instead, if we remove the vortex 
in the limit of the vanishing Josephson coupling,  
there remain bare (unconfined) instantons and monopoles 
inside the junction. 
While an instanton becomes a Skyrmion in the junction 
\cite{Eto:2005cc},
here we find that a monopole becomes a $U(1)^{N-1}$ vortex 
in the $U(N)$ principal chiral model  
with the twisted mass 
inside the junction, for which  
the monopole charges 
$\pi_2 [SU(N)/U(1)^{N-1}] \simeq {\mathbb Z}^{N-1}$ 
coincide with the vortex charges  
$\pi_1 [U(1)^{N-1}] \simeq {\mathbb Z}^{N-1}$.
We give an explicit ansatz for a single $U(1)$ vortex for $N=2$.
A quite nontrivial check is given 
by turning on 
the Josephson interaction in this configuration;
We find that there must appear two sine-Gordon solitons 
with opposite ${\mathbb C}P^1$ orientations 
attached to the vortex from its both sides. 
These sine-Gordon solitons are nothing but 
Josephson vortices, and so 
this configuration is precisely the case of a confined monopole.
For general $N$, we find $N-1$ vortices 
connected or attached by $N$ sine-Gordon solitons. 
As mentioned above, 
a T-duality between monopoles and instantons  
leads a T-duality between ${\mathbb C}P^{N-1}$ kinks
and ${\mathbb C}P^{N-1}$ lumps  
with a twisted boundary condition inside 
a non-Abelian vortex 
\cite{Eto:2004rz}.
Here, we find that 
$U(1)^{N-1}$ vortices are dimensionally reduced from (T-dual to) 
$SU(N)$ Skyrmions with a twisted boundary condition. 
The case of $N=2$ was found before in Ref.~\cite{Harland:2008eu},
in which numerical solutions were obtained. 
The ${\mathbb C}P^{N-1}$ lumps 
inside a non-Abelian sine-Gordon soliton 
in the $U(N)$ principal chiral  model 
are the $SU(N)$ Skyrmions \cite{Eto:2015uqa}.
Therefore, they are all instantons 
if we realize the $U(N)$ principal chiral  model 
inside the junction (non-Abelian domain wall). 
Thus, our configurations 
suggest an another  
duality between the ${\mathbb C}P^{N-1}$ lumps 
and the $SU(N)$ Skyrmions 
(both instantons in the bulk)
as well as that between the ${\mathbb C}P^{N-1}$ kinks and 
the $U(1)^{N-1}$ vortices 
(both monopoles in the bulk).
Since the former (latter) are both instantons (monopoles) 
viewed from different host solitons, 
the non-Abelian vortex on one hand 
and the non-Abelian domain wall on the other hand, 
this duality may be understood as 
T-dualities between these host solitons. 
All these relations are 
summarized in Fig.~\ref{fig:monopole-instanton}.

Instantons (or solitons) are fractionalized, 
that is, a single instanton (soliton) of unit topological charge 
is decomposed into 
multiple fractional instantons (solitons) 
with fractional topological charges  
in the presence of a twisted boundary condition. 
Fractional instantons in 
the ${\mathbb C}P^{N-1}$ model \cite{Eto:2004rz} 
(see also Refs.~\cite{Bruckmann:2007zh}) and 
the Grassmann sigma model
\cite{Eto:2006mz} in two Euclidean dimensions 
have been paid renewal interests 
because their composites with zero instanton charges, called bions, play significant roles in the resurgence 
of quantum field theory 
\cite{Dunne:2012ae,Dabrowski:2013kba,
Misumi:2014jua,Shermer:2014wxa}.
Fractional instantons and bions 
in the $O(N)$ model 
in Euclidean $N-1$ dimensions 
were also studied in Ref.~\cite{Nitta:2014vpa},
where the cases of $N=2$ and $3$ correspond to 
the ${\mathbb C}P^1$ model in two dimensions 
and 
the $SU(2)$ principal chiral model in three dimensions, respectively. 
Those in the principal chiral model in three dimensions were studied in Ref.~\cite{Nitta:2015tua}.
Our configuration studied in this paper 
suggests a kind of duality between 
fractional instantons and bions in these models, 
and for general $N$, those 
in the ${\mathbb C}P^{N-1}$ model in two dimensions 
and $SU(N)$ principal chiral model in three dimensions.

In a conventional Josephson junction of 
two metallic superconductors, 
electrons carry quantum tunneling.
Monopoles carry a quantum tunneling 
in a dual Josephson junction \cite{Dvali:2007nm}, 
that is a junction of two confinement phases (as dual superconductors) \cite{Tetradis:1999tn},
where quarks are confined and 
monopoles are considered to be condensed.
In contrast to this, our case corresponds to
unconfined monopoles stably existing 
inside a usual Josephson junction 
of two color superconductors, 
where quarks are condensed and monopoles are confined 
\cite{Eto:2011mk}.
Therefore, it suggests that, 
as dual to this, unconfined quarks can stably exist 
in a dual Josephson junction of two 
confinement phases.

This paper is organized as follows. 
After our model is given in Sec.~\ref{sec:model}, 
we summarize the non-Abelian Josephson 
junction and non-Abelian Josephson vortices
in Sec.~\ref{sec:SGkink}.
In Secs.~\ref{sec:instantons} and \ref{sec:monopoles}, 
we construct Josephson instantons and monopoles, 
respectively.
Section \ref{sec:summary} is devoted to a summary 
and discussion. 

\newpage

\section{The model \label{sec:model}}
The theory that we consider is 
a $U(N)$ gauge theory in the Higgs phase 
in $d=3+1$ (or $d=4+1$) dimensions 
with the following matter contents: 
a $U(N)$ gauge field $A_{\mu}$,  
two $N$ by $N$ charged complex scalar fields 
$H = (H_1,H_2)$,  
and a real adjoint  
$N$ by $N$ scalar field $\Sigma(x)$.  
The Lagrangian is given as follows:
\beq
&& {\cal L} \,=\, -\1{4 g^2} \tr F_{\mu\nu}F^{\mu\nu} 
 + \1{g^2} \tr (D_{\mu} \Sigma)^2\, 
 + \tr |D_{\mu} H|^2 + {\cal L}_J  - V
\eeq
where $V$ is the potential term
\beq
&& V \,=\, {g^2 \over 4} \tr ( H H^\dagger -v^2 {\bf 1}_N)^2 
 \,+ \tr |\Sigma H - H M|^2 ,
\eeq
and $D_{\mu}$ is the covariant derivative, 
given by  
$D_{\mu} H = \del_{\mu} H - i A_{\mu } H$ 
and 
$D_{\mu} \Sigma = \del_{\mu} \Sigma - i [A_{\mu},\Sigma]$, 
$g$ is the gauge coupling constant that we take 
common for 
the $U(1)$ and $SU(N)$ factors of $U(N)$, 
$v$ is a real constant representing the 
vacuum expectation value of $H$, and   
$M$ is a $2N$ by $2N$ mass matrix for $H$ 
given below.
Here, 
${\cal L}_J$ is a scalar coupling that we call 
the Josephson interaction 
\beq
&& {\cal L}_{J} \,=\, -
 \gamma \tr (H_1^\dagger H_2  + H_2^\dagger H_1) \,\,
\label{eq:Josephson}
\eeq
motivated by the Josephson junction of 
two superconductors.
Instead of this term,
we may consider a quadratic Josephson term 
${\cal L}_{J,2} \,=\, -
 \gamma \tr [(H_1^\dagger H_2)^2  + (H_2^\dagger H_1)^2]$
 \cite{Nitta:2014rxa} 
that is a non-Abelian extension of 
the Josephson term 
in chiral p-wave superconductors \cite{Agterberg1998}. 
Apart from the Josephson term $ {\cal L}_J$ 
(or $ {\cal L}_{J,2}$), 
the model is a truncation of the bosonic part of 
${\cal N}=2$ supersymmetric 
theory (with eight supercharges) 
in $d=3+1$ (or $4+1$) \cite{Eto:2006pg}. 

The $U(N)$ gauge (color) symmetry acts on fields as
\beq
\,
 A_{\mu} \to g A_{\mu} g^{-1} + i g \del_{\mu} g^{-1},  \quad
 H \to g H , \quad 
 \Sigma \to g \Sigma g^{-1}, \quad 
\, g \in U(N)_{\rm C},
\eeq
while the flavor (global) symmetry depends on the mass matrix;
In the massless case $M =0$, 
the flavor symmetry is the maximum $SU(2N)$.
This is explicitly broken by 
the mass matrix $M$ that we take 
\beq 
 M \,=\, {\rm diag.}(m {\bf 1}_N,- m {\bf 1}_N)
\eeq 
with a real constant $m$, together with 
a small mass perturbation
\beq 
 M={\rm diag.}(m {\bf 1}_N + \Delta M, 
- m {\bf 1}_N - \Delta M) , \quad
\Delta M = {\rm diag.} (m_1,m_2,\cdots,m_N)
\label{eq:mass}
\eeq
where real mass shifts $m_a$ are much smaller than $m$:
$m_a \ll m$.
For $m \neq 0$ with $\Delta M=0$, the flavor symmetry  is
 $SU(N)_{\rm L} \times  SU(N)_{\rm R} \times U(1)_{\rm L-R}$, given by 
\beq
 H_1 \to H_1 U_{\rm L} e^{+i \alpha}, \quad 
 H_2 \to H_2 U_{\rm R} e^{-i \alpha},  \quad 
 U_{\rm L,R} \in SU(N)_{\rm L,R}, \quad 
 e^{i\alpha} \in  U(1)_{\rm L-R} ,
\label{eq:flavor}
\eeq
while for $\Delta M \neq 0$ with 
non-degenerate mass perturbation 
$m_a \neq m_b$ for $a \neq b$, 
the flavor symmetry is explicitly broken to 
$U(1)^{N-1}_{\rm L} \times U(1)^{N-1}_{\rm R}\times U(1)_{\rm L-R}$:
\beq
 H_1 \to H_1 U_{\rm L} e^{+i \alpha}, \quad 
 H_2 \to H_2 U_{\rm R} e^{-i \alpha},  \quad U_{\rm L,R} \in 
U(1)^{N-1}_{\rm L,R} \subset SU(N)_{\rm L,R}.
\label{eq:flavor2}
\eeq
In this paper, we mostly consider this non-degenerate case.

The vacuum structures of the model are as follows. 
In the massless case $m=0$ and $\Delta M =0$, 
the vacuum can be taken without the lost of generality as
\beq
\, H 
\,= \,
\left(
 v {\bf 1}_N ,{\bf 0}_N 
\right) ,\,
\quad
\quad \Sigma ={\bf 0}_N\,\,
\eeq 
by using the $SU(2N)$ flavor symmetry.
The unbroken symmetry is 
$SU(N)_{\rm C+L} \times SU(N)_{\rm R} \times U(1)$, 
in which the factor $SU(N)_{\rm C+L}$ is 
the color-flavor locked (global) symmetry.
The moduli space of vacua is  
the complex Grassmann manifold 
\cite{Higashijima:1999ki} 
\beq  
\,\, Gr_{2N,N}
 \,\simeq\, {SU(2N) \over SU(N)\times SU(N)\times U(1)}.\,
\label{eq:Grassmann}
\eeq 

In the massive case, $m \neq 0$ but still $\Delta M =0$, 
the above vacua are split into the following two disjoint vacua  \cite{footnote1} 
\beq
&& H \,=\, 
\left(
 v {\bf 1}_N ,{\bf 0}_N 
\right) , \quad
\Sigma = + m {\bf 1}_N:  
 \quad SU(N)_{\rm C+L} ,\,
\non
&& H \,=\, 
\left(
  {\bf 0}_N , v{\bf 1}_N 
\right) ,  \quad \Sigma = - m {\bf 1}_N:
\quad SU(N)_{\rm C+R} \,
 \label{eq:vac}
\eeq 
with the unbroken color-flavor locked (global) symmetries
$g=U_{\rm L}$ and $g = U_{\rm R}$, respectively.
These vacua are
color-flavor locked vacua  
that can be interpreted as 
non-Abelian color superconductors.  

With the non-degenerate mass deformation 
$\Delta M \neq 0$, each vacuum 
in Eq.~(\ref{eq:vac}) is shifted to
\beq
&& H = 
\left(
 v' {\bf 1}_N ,{\bf 0}_N 
\right) , \quad
\Sigma =  + m {\bf 1}_N + \Delta M:  
 \quad U(1)^{N-1}_{\rm C+L} ,
\non
&& H = 
\left(
  {\bf 0}_N , v' {\bf 1}_N 
\right) ,  
\quad \Sigma = - m {\bf 1}_N - \Delta M:
\quad U(1)^{N-1}_{\rm C+R},
 \label{eq:vac2}
\eeq 
where $v'$ is shifted from $v$.

In the following sections, 
we often work in the strong  coupling 
(nonlinear sigma model) limit 
$g \to \infty$ 
for explicit calculations.  
In this limit, we have the constraints 
\beq
&&\,
H H^\dagger = v^2 {\bf 1}_N,\,\\
&&\, 
\Sigma = {H M H^\dagger \over H H^\dagger}
= v^{-2} H M H^\dagger,\,\label{eq:Sigma}\\
&&\, 
A_{\mu} 
= {i\over 2} v^{-2} [H \del_{\mu} H^\dagger -  (\del_{\mu}H)  H^\dagger], \, \label{eq:Amu}
\eeq
and the model is reduced 
to the Grassmann sigma model 
with the target space given in Eq.~(\ref{eq:Grassmann})
together with a potential term, known as the massive (twisted-mass deformed) Grassmann sigma model \cite{Arai:2003tc}.

\section{Non-Abelian Josephson junction 
and non-Abelian Josephson vortex
\label{sec:SGkink}}

\subsection{Non-Abelian Josephson junction 
as a non-Abelian domain wall}
For a while, we consider the case 
in the absence of the Josephson term $\gamma=0$ 
and the mass deformation $\Delta M=0$, 
and then we turn them on later. 
In the sigma model limit,
a non-Abelian domain wall solution 
interpolating between the two vacua 
in Eq.~(\ref{eq:vac}) can be obtained as 
\cite{Isozumi:2004jc,Shifman:2003uh,
Eto:2005cc,Eto:2008dm}
\beq
&&\, H_{\rm wall,0} = {v \over \sqrt {1+|u_{\rm wall}|^2}}
      \left({\bf 1}_N, u_{\rm wall} {\bf 1}_N\right),\,
\quad \,
u_{\rm wall}(x^1) = e^{\mp m (x^1-X^1) + i \ph} ,\, 
  \label{eq:wall-sol0}
\eeq
with $\Sigma$ and $A_1$ obtained from 
Eqs.~(\ref{eq:Sigma}) and (\ref{eq:Amu}),
where we place it perpendicular to the $x^1$ coordinate, 
 and $X^1$ is its position in that coordinate 
or the translational modulus. 
$u_{\rm wall}$ is a domain wall solution in 
the massive ${\mathbb C}P^1$ model \cite{Abraham:1992vb}
with width $m^{-1}$. 
The most general solution can be obtained from 
the solution in Eq.~(\ref{eq:wall-sol0})  
 by acting the $SU(N)_{\rm C+L+R}$ symmetry that remains in the vacuum on the above solution: 
\beq 
&& \,H_{\rm wall} = V H_{\rm wall,0} 
\left(\begin{array}{cc}
 V^\dagger & 0 \\ 0 & V
\end{array} \right) 
\,=  {v\over \sqrt {1+ e^{\mp 2 m (x^1-X^1) }}}
      \left({\bf 1}_N, e^{\mp m (x^1-X^1) }U \right), 
\eeq 
with $V \in SU(N)$. 
Here we have defined the group-valued moduli $U$ by 
$U \equiv V^2 e^{i\ph} \in U(N)$.
This transformation gives the domain wall  
the moduli $U \in U(N)$ in addition to $X^1$ \cite{footnote2}
\beq 
\, (X^1,U) \in  {\cal M}_{\rm wall} \,\simeq \, {\mathbb R} \times U(N).
\, \label{eq:wall-moduli}
\eeq 

Now 
we turn on the Josephson interaction $\gamma$ 
so that the domain wall becomes the Josephson junction. 
The effective theory of the non-Abelian 
Josephson junction can be 
constructed by using the (Manton's) moduli approximation \cite{Manton:1981mp,Eto:2006uw};
First, we promote the moduli parameters $X^1$ and $U$ to moduli fields 
 $X^1(x^i)$ and $U(x^i)$, respectively ($i=0,2,3,(4)$ for $d=3+1\, (4+1)$) 
on the world volume of the domain wall,  
and then perform integration over the codimension. 
We thus obtain the effective theory given by \cite{Shifman:2003uh,Eto:2005cc,Eto:2008dm}:
\beq
 \,{\cal L}_{\rm wall} =  
{v^2 \over 2m} \del_i X^1 \del^i X^1 
-  f_{\pi}^2 
\tr \left(U^\dagger \del_{i} U
            U^\dagger \del^{i} U \right) 
+  {\cal L}_{{\rm wall},J} ,\,
\quad  f_{\pi}^2 \equiv {v^2 \over 4m} 
\label{eq:eff}
\eeq
with the mass term induced from 
the non-Abelian Josephson term \cite{Nitta:2015mma}
\beq
 {\cal L}_{{\rm wall},J}
= - m'^2  (\tr U + \tr U^\dagger) ,\quad
\label{eq:eff-Josephson}
m'^2 \equiv  {\pi \gamma \over 2m}.
\eeq
We thus obtain 
the $U(N)$ chiral Lagrangian in Eq.~(\ref{eq:eff}), 
and the term in Eq.~(\ref{eq:eff-Josephson}) 
is nothing but the conventional pion mass term. 
This potential term lifts 
the $U(N)$ vacuum manifold,  
leaving the unique vacuum
\beq
 U = {\bf 1}_N ,
\eeq
as the case of the usual chiral Lagrangian.

\subsection{Non-Abelian vortex as a non-Abelian sine-Gordon soliton inside the junction}

In this subsection, we discuss a non-Abelian vortex.
First, we discuss a non-Abelian vortex in the absence of 
the Josephson junction, and later consider them together.
The non-Abelian vortices in the massless case 
$m=0$ and $\Delta M=0$ are non-Abelian 
semi-local vortices \cite{Shifman:2006kd},
but in the massive case they become local vortices 
of the Abrikosov-Nielsen-Olesen (ANO) 
vortex type \cite{Abrikosov:1956sx}.
In the left (right) vacuum in Eq.~(\ref{eq:vac}), 
we can neglect $H_2$ ($H_1$).
There, the $U(N)$ symmetry is spontaneously broken completely, 
locking with the $SU(N)_{\rm L(R)}$ flavor symmetry 
to the $SU(N)_{\rm C+L(R)}$ color-flavor locked symmetry.
A non-Abelian vortex solution  in the $x^1$-$x^2$ plane 
with a non-Abelian magnetic field $F_{12}$
and a scalar field, is given by 
\beq
 && F_{12,0} = \, {\rm diag} (F_*(r),0,\cdots,0), \quad
 (H_{1(2)})_0 = v \, {\rm diag} (f(r)e^{i \theta},1,\cdots,1),
 \label{eq:NA-vortex}
\eeq
with the boundary conditions for the profile function $g$,    
$g(r) \to 1$ ($r\to \infty$) and 
$g(r) \to 0$ ($r=0$).
Here $(r,\theta)$ are the polar coordinates in 
the $x^1$-$x^2$ plane.  
This solution is obtained by  
embedding of the ANO vortex solution \cite{Abrikosov:1956sx} 
$(F_*(r), g(r)e^{i \theta})$ into the upper-left corner. 
The most general solution can be obtained by acting the 
color-flavor locked symmetry $SU(N)_{\rm C+L(R)}$ to 
the above solution: 
\beq
 && F_{12} = V {\rm diag} (F_*(r),0,\cdots,0) V^\dagger, \quad
  H_{1(2)} = v \, V{\rm diag} (f(r)e^{i \theta},1,\cdots,1) V^\dagger, \non
&& V \in SU(N).
 \label{eq:NA-vortex}
\eeq
This solution spontaneously breaks 
the color-flavor locked symmetry $SU(N)_{\rm C+L(R)}$ 
into a subgroup 
$SU(N-1)\times U(1)$. Therefore, it results in 
the moduli localized on the vortex core;
\beq
\,{\cal M}_{\rm vortex} 
\,\simeq \,  {\mathbb C} \times {\mathbb C}P^{N-1} 
= {\mathbb C} \times {SU(N)_{\rm C+L(R)} \over SU(N-1)\times U(1)}, \,
\label{eq:vortex-moduli}
\eeq
which are called the orientational moduli.

\begin{figure}[ht]
\begin{center}
\includegraphics[width=0.5\linewidth,keepaspectratio]{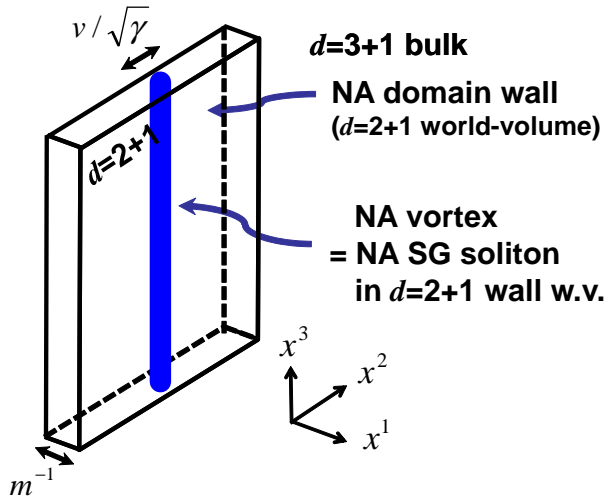}
\caption{A schematic picture of 
a non-Abelian sine-Gordon soliton in 
a non-Abelian domain wall describing 
a non-Abelian vortex. 
\label{fig:SG}}
\end{center}
\end{figure}
When the non-Abelian vortex is placed parallel to 
the non-Abelian Josephson junction (domain wall), 
it is absorbed into the junction 
to minimize the total energy (see Fig.~\ref{fig:SG}).  
The resulting configuration can be described as 
a non-Abelian sine-Gordon soliton
in the $U(N)$ chiral Lagrangian in Eq.~(\ref{eq:eff}) 
with the mass term in Eq.~(\ref{eq:eff-Josephson}). 
A non-Abelian sine-Gordon soliton 
(perpendicular to the $x^2$ coordinate) is given as  
\cite{Nitta:2014rxa,Eto:2015uqa}:
\beq
&& \,U(x) \,=\, {\rm diag}\, (u(x^2),1,\cdots,1) , \, \label{eq:embedding}\\
&& \,u(x^2) \,=\, \exp i \theta_{\rm SG}(x^2) 
= \exp \left(4 i \, \arctan \exp [m''  (x^2- X^2)] \right) ,\,
\label{eq:U(1)-one-kink}
\eeq
with the translational modulus $X^2$ and the effective mass $m''$ defined by 
\beq
 m''^2 = \frac{m'^2}{f_\pi^2} =\frac{2\pi \gamma}{v^2}.\label{eq:m''}
\eeq
The width of the soliton is $m''^{-1} \sim v/\sqrt{\gamma}$, 
and the tension of the soliton is 
\beq 
T_{\rm SG} = 8m''. \label{eq:SG-tension}
\eeq
The most general single soliton solution can be obtained 
by acting the $SU(N)$ symmetry on Eq.~(\ref{eq:U(1)-one-kink}):
\beq
 U(x) = V{\rm diag} (u(x^2),1,\cdots,1) V^\dagger , 
  \quad V\in SU(N).  \label{eq:SU(N)moduli}
\eeq
Therefore, the non-Abelian sine-Gordon soliton carries 
the orientational moduli  
\beq
 {\cal M}_{\rm NA SG}  \simeq  {\mathbb R}\times {\mathbb C}P^{N-1} \simeq {\mathbb R}\times {SU(N) \over SU(N-1) \times U(1)},
\eeq
that coincides with the moduli of the non-Abelian vortex  
in Eq.~(\ref{eq:vortex-moduli}), except for 
one translation modulus $X^1$ transverse to the junction.
The composite configuration (in the $x^1$-$x^2$ plane) can be written as 
\beq 
H_{\rm composite} 
\,=\,  \1{\sqrt {1+ e^{\mp 2 m (x^1-X^1) }}}
      \left({\bf 1}_N, e^{\mp m (x^1-X^1) }V
\,{\rm diag} \,
(e^{i \theta_{\rm SG}(x^2)},1,\cdots,1) V^\dagger \right) .\,
\label{eq:total}
\eeq
It was shown in Ref.~\cite{Nitta:2015mma} 
from the flux matching that this is precisely a non-Abelian vortex. 
The coordinates of the configurations were summarized 
in (the first two lines of) Table \ref{table:config}.
This composite configuration is non-BPS. 
In fact, the Josephson term stabilizing the vortex cannot be made 
supersymmetric. 
\begin{table}
\begin{tabular}{|c|ccccc|}
\hline
                        & $x^0$ & $x^1$ & $x^2$ & $x^3$ & $x^4$ \\ \hline
NA wall (Josephson junc) &  $\circ$ & $\times$ & $\circ$ & $\circ$  & $\circ$\\
NA vortex         & $\circ$ & $\times$ & $\times$ & $\circ$ & $\circ$\\
monopole         & $\circ$ &  $\times$ &  $\times$ & $\times$ & $\circ$\\
YM instanton    & $\circ$ &  $\times$ &  $\times$ & $\times$ & $\times$ \\ \hline
\end{tabular}
\caption{
The space-time configurations of the topological solitons in this paper. 
Here,
``$\times$" denote the codimensions that soliton configurations depend on, 
while ``$\circ$" denote the world-volume directions that 
the static soliton configurations do not depend and the moduli fields live in.
\label{table:config}}
\end{table}

The effective theory of the sine-Gordon soliton with the world-volume 
$x^{\alpha}$ ($\alpha=0,3,(4)$ for $d=3+1\, (4+1)$) can 
be also obtained by the moduli approximation \cite{Eto:2015uqa}:
\beq
{\cal L}_{\rm SG} &=&
C_X
\p_\alpha X^2 \p^\alpha X^2 
+ C_{\phi} 
\left[
\p_\alpha \phi^\dagger \p^\alpha \phi + (\phi^\dagger\p_\alpha \phi)(\phi^\dagger\p^\alpha \phi)
\right]
\eeq
with the constants (called K\"ahler classes)  
\beq
&& C_X =  \frac{f_\pi^2 T_{\rm SG}}{2} = \sqrt{2\pi} \frac{v{\sqrt \gamma}}{m},
\quad
C_{\phi} = \frac{f_\pi^2 T_{\rm SG}}{m''^2} = \sqrt{2\over \pi}{v^3 \over m \sqrt \gamma} .
\eeq
Here, the first equalities were derived in Ref.~\cite{Eto:2015uqa} 
and the second equalities hold from Eqs.~(\ref{eq:m''}) and (\ref{eq:SG-tension}).

\section{Instantons inside a non-Abelian Josephson junction:
 Josephson instantons
\label{sec:instantons}}

When we study Yang-Mills instantons, 
we promote the dimensionality of space-time to 
$d= 4+1$, and consider instanton-particles in 
four Euclidean space in $d= 4+1$ dimensions.

\begin{figure}[ht]
\begin{center}
\includegraphics[width=0.7\linewidth,keepaspectratio]{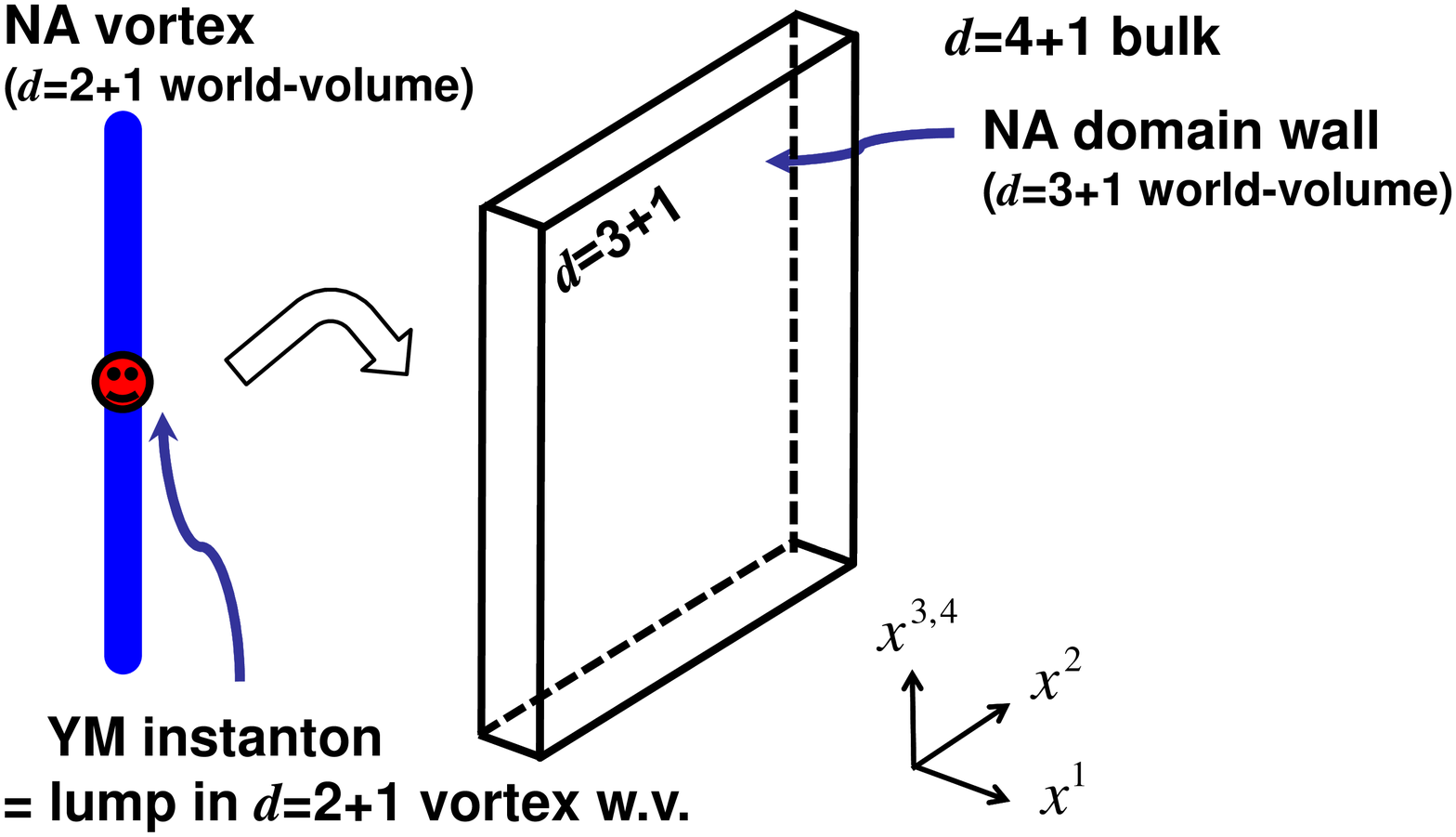}\\
(a)\\
\includegraphics[width=0.6\linewidth,keepaspectratio]{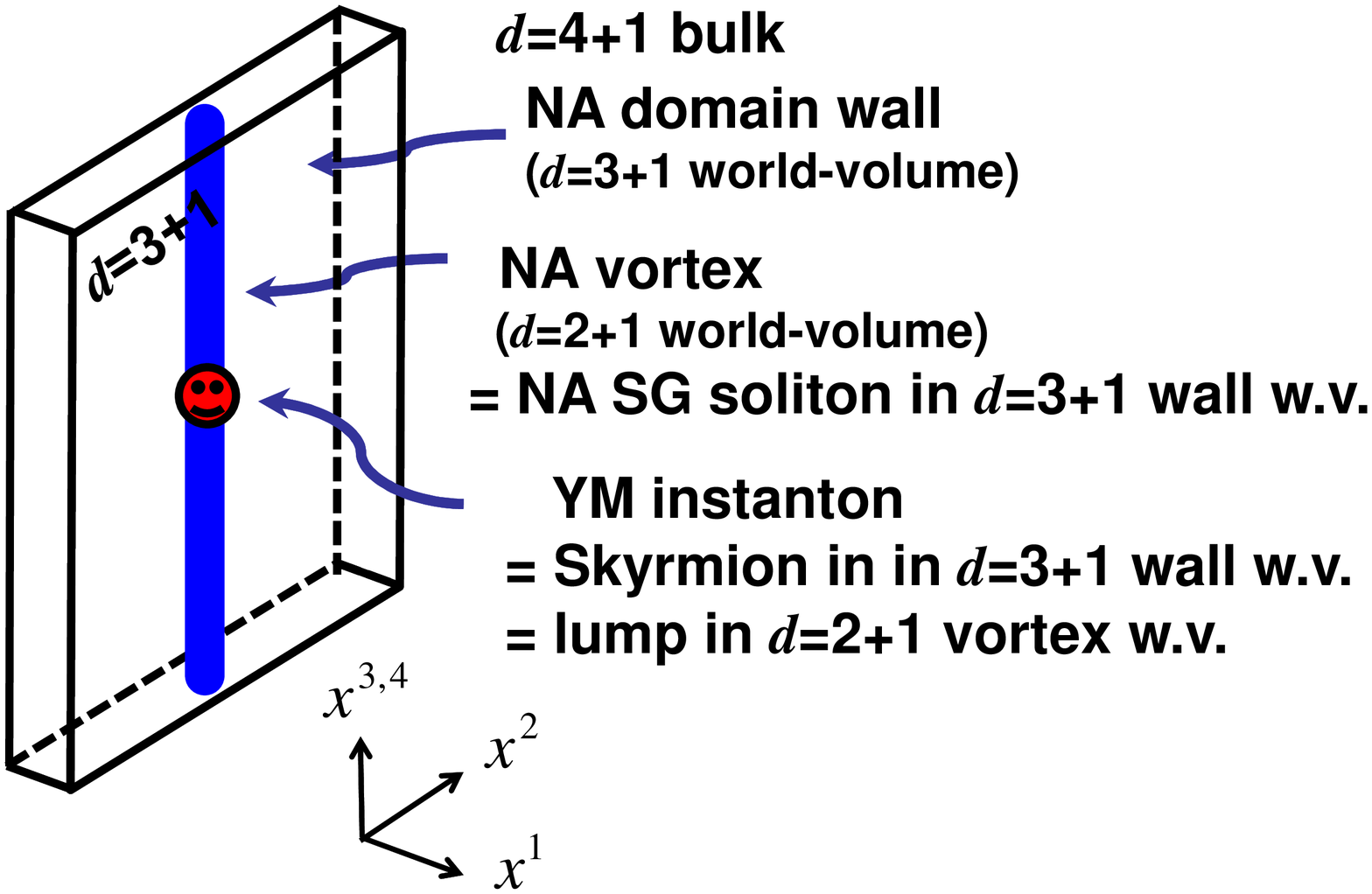}\\
(b)
\caption{A Yang-Mills instanton as a lump inside a non-Abelian vortex (a) apart from (b) inside a junction (domain wall).
\label{fig:instanton}}
\end{center}
\end{figure}

First we consider an instanton with the help of a non-Abelian vortex 
far apart from the junction 
as illustrated in Fig.~\ref{fig:instanton}(a). 
The non-Abelian vortex has the moduli in Eq.~(\ref{eq:vortex-moduli}). 
The effective theory of the non-Abelian vortex 
placed in the $x^1$-$x^2$ plane 
is therefore the ${\mathbb C}P^{N-1}$ model 
($\alpha=0,3,4$ for $d=4+1$)
\cite{Hanany:2004ea,Auzzi:2003fs,Shifman:2004dr,Eto:2004rz}
\beq
{\cal L}_{\rm vortex} = 
2\pi v^2 \p_\alpha Z \p^\alpha Z + {4\pi \over g^2}
\left[
\p_\alpha \phi^\dagger \p^\alpha \phi + (\phi^\dagger\p_\alpha \phi)(\phi^\dagger\p^\alpha \phi)
\right],
\eeq
with the complex position moduli $Z\equiv X^1+iX^2$ of the vortex and 
a complex $N$-vector $\phi$ with a constraint 
$\phi^\dagger \phi=1$.
The ${\mathbb C}P^{N-1}$ model admits 
${\mathbb C}P^{N-1}$ lumps 
classified by $\pi_2 ({\mathbb C}P^{N-1})\simeq {\mathbb Z}$. 
The ${\mathbb C}P^{N-1}$ lumps (in the $x^3$-$x^4$ plane)
in the vortex effective theory 
can be identified with Yang-Mills instantons in the bulk \cite{Eto:2004rz}. 
This can be verified from the lump energy $E_{\rm lump} $, 
coinciding with the instanton energy $E_{\rm inst}$ \cite{Eto:2004rz}:
\beq 
E_{\rm lump} = {4\pi \over g^2} T_{\rm lump} 
= {4\pi \over g^2} \times 2\pi k 
= {8\pi^2 \over g^2} k = E_{\rm inst}.
\eeq 
Here, $T_{\rm lump}  = 2\pi k$ is the lump charge with 
the lump number $k \in \pi_2 ({\mathbb C}P^{N-1})\simeq {\mathbb Z}$.

In the presence of the non-Abelian Josephson junction,
this composite soliton, 
a vortex-instanton composite, will be absorbed into 
the junction to minimize the total energy.
We then obtain an instanton inside the non-Abelian vortex inside the junction 
as illustrated in Fig.~\ref{fig:instanton}(b).
The non-Abelian vortex in the bulk is  
a non-Abelian sine-Gordon soliton inside 
the junction whose effective theory is 
the $U(N)$ principal chiral model. 
It was further shown in  Ref.~\cite{Eto:2015uqa} that  
${\mathbb C}P^{N-1}$ 
lumps inside the non-Abelian sine-Gordon soliton 
are $SU(N)$ Skyrmion in the $U(N)$ principal chiral model  as follows:
The baryon (Skyrmion) number 
$B$ taking a value in $\pi_3 [SU(N)]\simeq {\mathbb Z}$ 
in the bulk can be calculated as $(i=2,3,4)$
\beq
B &=& \1{24 \pi^2} \int d^3x\  \epsilon_{ijk} 
\tr (U^\dagger \del_i U  U^\dagger \del_j U U^\dagger \del_k U) \non
&=& -\1{8 \pi^2} \int  d^3x\  \tr \left[\left(\p_3U^\dagger\p_4U - \p_4U^\dagger\p_3U\right)  U^\dagger \p_2 U \right] \non
&=&  \frac{i}{2\pi}\int dzd\bar z\ \tr\left(
\left[\p_{\bar z} {\cal P},\p_z{\cal P}\right] {\cal P}
\right) \times \frac{1}{2\pi} \int dx^2\  (1-\cos\theta_{\rm SG}) \p_2\theta_{\rm SG}\non
&=& k l
\eeq
with the projector 
${\cal P} = \phi \phi^\dagger$, $z\equiv x^3+i x^4$,   
the lump number $k \in \pi_2 ({\mathbb C}P^{N-1}) 
\simeq {\mathbb Z}$, 
and 
the sine-Gordon soliton number $l \in \pi_1[U(1)] \simeq {\mathbb Z}$ defined by
\beq
 l \equiv {\theta_{\rm SG}(x^2=+\infty) - \theta_{\rm SG}(x^2=-\infty) \over 2\pi} .
\eeq
Therefore, the ${\mathbb C}P^{N-1}$ lumps
on the non-Abelian sine-Gordon soliton are 
$SU(N)$ Skyrmions in the principal chiral model in 
$d=3+1$ dimensions.  
This composite configurations is non-BPS. 

Finally,  by following Ref.~\cite{Eto:2005cc} we discuss that  
the $SU(N)$ Skyrmions inside the non-Abelian Josephson junction (domain wall) can be identified 
with Yang-Mills instanton particles in the $d=4+1$ bulk.
Far apart from the domain wall, 
the gauge field $A_{\mu}$ falls into 
a pure gauge $A_{\mu} = -i(\del_{\mu} U) U^\dagger$.
In this setting, 
the instanton number coincides with the baryon number:
\beq
 I &=& {1\over 24 \pi^2} \int d^4 x \epsilon^{\mu\nu\rho\sigma}F_{\mu\nu}F_{\rho\sigma}  
   =  {1\over 24 \pi^2} 
\int_{{\mathbb R}^3({x^1= +\infty}) 
    - {\mathbb R}^3({x^1= -\infty}) }   d^3 x 
\epsilon^{ijk}
\tr (U^\dagger \del_iU U^\dagger \del_jU U^\dagger \del_kU )
 \non
&=& \int_{{\mathbb R}^3({x^1= +\infty})} d^3x
\epsilon^{ijk}
\tr (U^\dagger \del_iU U^\dagger \del_jU U^\dagger \del_kU )  
 = B.
\eeq
Therefore, we have a consistent picture.
The $SU(N)$ Yang-Mills instantons are $SU(N)$ Skyrmions in the non-Abelian domain wall (Josephson junction) and are 
${\mathbb C}P^{N-1}$ lumps inside the non-Abelian vortex
as illustrated in Fig.~\ref{fig:instanton}(b).

We have two limits to remove host solitons 
as illustrated in Fig.~\ref{fig:monopole-instanton}
(a), (b) and (c), 
where we have drawn configurations inside 
the Josephson junction (wall), ignoring the outside. 
The two limits are:
(1) $m\to 0$: 
the non-Abelian domain wall disappears in this limit,  
since its width  is $m^{-1}$.
The configuration is a vortex-instanton composite,
where instantons are lumps in the vortex. 
This composite is 1/4 BPS if embedded into a 
supersymmetric theory \cite{Eto:2004rz}. 
\\
(2) $\gamma \to 0$: 
the non-Abelian vortex disappears in this limit, 
since the size of the vortex along the domain wall 
world-volume is proportional to $m''^{-1} \sim v/\sqrt{\gamma}$.
The configuration is a wall-instanton composite,
where instantons are Skyrmions in the wall. 
This composite is non-BPS \cite{Eto:2005sw} 
even in the absence of 
the Josephson term which cannot be made supersymmetric. 

Therefore, the original configuration 
gives a kind of duality between 
${\mathbb C}P^{N-1}$ lumps 
and $SU(N)$ Skyrmions 
both realized by Yang-Mills instantons.

\section{Monopoles inside a non-Abelian Josephson junction: 
Josephson monopoles
\label{sec:monopoles}}

\begin{table}
\begin{tabular}{c|c|c}
bulk         & NA vortex & NA wall (Josephson junc) \\ \hline
$SU(N)$ instanton  & ${\mathbb C}P^{N-1}$ lump & 
$SU(N)$ Skyrmion \\
$\Downarrow$ SS red & $\Downarrow$ SS red & $\Downarrow$ SS red\\
$SU(N)/U(1)^{N-1}$ monopole & ${\mathbb C}P^{N-1}$ kink & 
$U(1)^{N-1}$ vortex
\end{tabular}
\caption{
T-duality relations (Scherk-Schwartz dimensional reductions). 
$SU(N)$ instantons are dimensionally reduced to 
$SU(N)/U(1)^{N-1}$ (Abelian) monopoles in the bulk 
for non-degenerate twisted masses. 
This relation leads two duality relations; 
${\mathbb C}P^{N-1}$ lumps are dimensionally reduced to 
${\mathbb C}P^{N-1}$ kinks inside the non-Abelian vortex, while 
$SU(N)$ Skyrmions are dimensionally reduced to the $U(1)^{N-1}$ 
vortices inside the non-Abelian domain wall or Josephson junction.
\label{table:SS-dim-red}}
\end{table}

In this section, we discuss monopoles in the non-Abelian 
Josephson junction.
We now turn on the mass perturbation $\Delta M$ 
in Eq.~(\ref{eq:mass}).
It is useful to note that 
this mass can be obtained 
from the theory with $\Delta M=0$ 
with the compactified $x^4$ direction with 
the twisted boundary condition 
along the $x^4$ coordinate:
\beq 
 H (x^{\mu},x^4+R) 
=  H (x^{\mu},x^4) 
\left(
\begin{array}{cc}
\exp (i \Delta M) & {\bf 0}_N \\
                  {\bf 0}_N & \exp (-i \Delta M)
\end{array}
\right)  \label{eq:twist}
\eeq 
with ${\mu}=0,1,2,3$. 
This twisting group element belongs to 
$U(1)^{N-1}$ in the $SU(N)$ flavor symmetry.
By assuming the $x^4$ dependence of the fields as
\beq 
 H (x^{\mu},x^4) 
=  H (x^{\mu}) 
\left(
\begin{array}{cc}
\exp [i (x^4/R) \Delta M] & {\bf 0}_N \\
                  {\bf 0}_N & \exp [-i (x^4/R) \Delta M]
\end{array}
\right),  \label{eq:twist2}
\eeq 
we substitute this to the kinetic term, 
to obtain the mass deformation $\Delta M$ 
(we set $R=1$):
\beq
 \del_4 H (x^{\mu},x^4) 
= i (H_1 (x^{\mu}) \Delta M, -  H_2 (x^{\mu}) \Delta M) .
\eeq
With putting $A_4(x^{\mu},x^4) = \Sigma(x^{\mu})$
\beq
 \tr |D_4 H (x^{\mu},x^4)|^2 
= \tr |\Sigma H_1 - H_1 (x^{\mu}) \Delta M|^2 
+ \tr |\Sigma H_2 + H_2 (x^{\mu}) \Delta M|^2 .
\eeq

This is known as the Scherk-Schwarz dimensional reduction.
The Scherk-Schwarz dimensional reduction induces 
the twisted mass on the soliton world-volumes too.

\subsection{Monopoles inside a non-Abelian vortex}
The Scherk-Schwarz dimensional reduction 
acts on the moduli fields on a non-Abelian vortex as
\beq 
\phi(x^\alpha,x^4) = \exp [i (x^4/R) \Delta M] \phi(x^\alpha)
\label{eq:SSred-on-vortex}
\eeq
with the vortex world-volume coordinates 
$x^{\alpha}$ ($\alpha=0,3$). 
We then obtain 
\beq
{\cal L}_{{\rm vortex}, \Delta M} &=& 
2\pi v^2 \p_\alpha Z \p^\alpha Z + {4\pi \over g^2}
\left[
\p_\alpha \phi^\dagger \p^\alpha \phi + (\phi^\dagger\p_\alpha \phi)(\phi^\dagger\p^\alpha \phi)
\right] 
- V\non
V &=&   {4\pi \over g^2} 
\left[ (\phi^\dagger \Delta M \phi)^2- \phi^\dagger (\Delta M)^2 \phi \right] ,\label{eq:vortex-eff}
\eeq
that is known as the massive (or twisted-mass deformed) 
${\mathbb C}P^{N-1}$ model.
For non-degenerate mass deformation $\Delta M$,  
this potential admits $N$ discrete  
vacua
\beq
 \phi_a^T = (0,\cdots,0,1,0,\cdots) ,  \quad a=1,\cdots,N
\eeq
where only the $a$-component is nonzero.
The Lagrangian (\ref{eq:vortex-eff})
admits $N-1$ multi-kink solutions, 
where 
the constituent kink connecting the 
$a$-th and $a+1$-th vacua 
has the mass 
$E_{{\rm kink},a}$ ($a=1,\cdots,N-1$) that coincides with 
the mass of a monopole $E_{{\rm monopole},a}$ 
\cite{Tong:2003pz,Shifman:2004dr,Hanany:2004ea,Nitta:2010nd}:
\beq 
 E_{{\rm kink},a} ={4\pi \over g^2} (m_{a+1} - m_a) 
 =  E_{{\rm monopole},a}.
\eeq
This coincidence implies the coincidence of topological charges, since they are both BPS. 
The monopole-vortex composite configuration is illustrated in 
Fig.~\ref{fig:monopole-instanton}(d).

In the presence of the Josephson junction, this monopole-vortex composite is absorbed into it, 
resulting in a wall-vortex-monopole composite 
as in Fig.~\ref{fig:monopole-instanton}(e), 
that is discussed in the following subsections.

\subsection{Monopoles inside a non-Abelian domain wall}

Let us consider the case of $\gamma=0$ 
in this subsection, 
in which case we do not have any vortex.
We work in the domain-wall effective theory.  
The Scherk-Schwarz dimensional reduction 
along the compactified direction $x^4$
acts on the moduli fields $U$ 
on a non-Abelian domain wall 
as
\beq 
U(x^i,x^4) 
= \exp [i (x^4/R) \Delta M] U(x^i) 
   \exp [-i (x^4/R) \Delta M] 
\eeq
with $i=0,2,3$. 
Then, the derivative of the moduli 
with respect to the compactified direction is 
obtained as
\beq 
\del_4 U(x^i ,x^4) 
= i [ \Delta M, U(x^i ,x^4) ],
\eeq
and the gradient term in the $x^4$ direction  
can be calculated as
\beq 
\tr (i U^\dagger (x^i,x^4)  \del_4 U(x^i,x^4) )^2
= - \tr ([ \Delta M, U(x^i) ]^\dagger[ \Delta M, U(x^i) ]).
\eeq
We thus obtain the domain-wall effective theory 
(or the effective theory of the Josephson junction)
in the presence of the twisted mass $\Delta M$ 
in the original theory: 
\beq
&& {\cal L}_{{\rm wall},\Delta M} =  
{v^2 \over 2m} \del_i X \del^i X 
- {v^2 \over 4m} 
\tr \left(U^\dagger \del_{i} U
            U^\dagger \del^{i} U \right) - V\non
&& 
V = 
   {v^2 \over 4m}  \tr ([ \Delta M, U ]^\dagger[ \Delta M, U ])
\label{eq:eff2}
\eeq

The vacua of the domain-wall effective theory are given by 
the condition 
\beq
 [ \Delta M, U ]=0.
\eeq
When $\Delta M$ is non-degenerate,  
Eq.~(\ref{eq:mass}) 
with $m_a \neq m_b$ for $a \neq b$,
the moduli space $M$ of vacua is 
\beq 
 U = {\rm diag}\, 
(e^{i \alpha_1}, \cdots,e^{i \alpha_N}) :  \quad M \simeq U(1)^{N-1} 
\eeq
with $\sum_{a=1}^N \alpha_a=0$.
It has the nontrivial first homotopy group  
\beq
 \pi_1 (M) \simeq {\mathbb Z}^{N-1}
\eeq
admitting $N-1$ kinds of vortices.
These $N-1$ kinds of vortices 
correspond to the monopole charge 
$\pi_2 [SU(N)/U(1)^{N-1}] \simeq 
\pi_1 [U(1)^{N-1}] 
\simeq {\mathbb Z}^{N-1}$ 
and to $N-1$ kinks in the mass deformed 
${\mathbb C}P^{N-1}$ model.

To be more specific, 
let us consider the simplest case of $N=2$. 
In this case, the target space of the principal chiral model 
is $SU(2) \simeq S^3 \simeq O(4)/O(3)$ 
(except for the $U(1)$ part), 
that is the $O(4)$ model. 
Let us express the field $U$ in
terms of four reals scalar fields $n_A(x)$ ($A=1,2,3,4$)
with the constraint $\sum_A n_A^2 =1$:
\beq
U = i \sum_{a=1}^3 n_a \sigma^a + n_4 \mathbf{1}_2 ,
\eeq
where  $\sigma^a$ are the Pauli matrices and
$U^\dag U = \mathbf{1}_2$ is equivalent to
$\mathbf{n}\cdot\mathbf{n}=1$. The Lagrangian is 
\beq
 {\cal L} = 
{v^2 \over 4m}\del_i{\bf n}\cdot \del^i{\bf n} 
\eeq
The twisted boundary condition 
\beq 
  U(x^i,x^4) =  
e^{i (x^4/R) m_1 \sigma_3} U(x^i)  e^{-i (x^4/R) m_1 \sigma_3}
 \label{eq:tbcO4--++}
\eeq
can be rewritten as
\beq 
(n_1,n_2,n_3,n_4)(x^i,x^4) 
= \left(\hat n_1(x^i)\cos {m_1 \over R} x^4, 
\hat n_2(x^i) \sin {m_1 \over R} x^4 ,
\hat n_3(x^i), \hat n_4(x^i)\right),
\eeq
and the induced potential term is 
(we take $R=1$)
\beq 
 V_m  = {v^2 \over 4m} \int_0^R d x^4 \left[(\del_4 n_1)^2 + (\del_4 n_2)^2 \right] 
 = {v^2 m_1^2 \over 16m}   (\hat n_1^2 + \hat n_2^2) 
 = {v^2 m_1^2 \over 16m}  (1-\hat n_3^2 - \hat n_4^2) . 
\label{eq:SS-dim-red}
\eeq 
Numerical solutions in
the case of $N=2$ were constructed before in Ref.~\cite{Harland:2008eu}.

The vacuum condition $n_1=n_2=0$ gives 
the vacuum manifold 
$M \simeq S^1$: $n_3^2 + n_4^2=1$. 
Therefore, the first homotopy group 
$\pi_1(M) \simeq {\mathbb Z}$ 
admits one kind of a vortex.
A vortex solution is of the form:
\beq
&& n_3 + i n_4 = \cos f(r) e^{i \theta}, \quad
 n_1 + i n_2 = \sin f(r) e^{i \alpha},  
\label{eq:skyrme-vortex}\\
&& U = 
\left(\begin{array}{cc} 
    \cos f(r) e^{i \theta} &  -\sin f(r)  e^{-i\alpha} \\
    \sin f(r)  e^{i\alpha} &   \cos f(r) e^{-i \theta} 
\end{array}
\right), \label{eq:skyrme-vortex2}
\eeq
where $f$ is a profile function satisfying 
the boundary conditions
\beq
  f \to \pi/2  
\; {\rm for}\; r\to \infty, \quad
  f =  0
\; {\rm for}\; r=0, \label{eq:bc-f}
\eeq
where $(r,\theta)$ are polar coordinates in 
the $x^2$-$x^3$ plane.  
Here, $\alpha$ in Eq.~(\ref{eq:skyrme-vortex}) is 
a real constant representing 
a $U(1)$ modulus of the vortex. 
This vortex inside the non-Abelian domain wall (Josephson junction) is a monopole in the bulk,
see Fig.~\ref{fig:monopole-instanton}(f).
In fact, they have the same $U(1)$ moduli.

For general $N$, a vortex solution corresponding to 
the $a$-th monopole ($a$-th kink) 
can be obtained by embedding the $N=2$ solution 
in Eq.~(\ref{eq:skyrme-vortex2}) to $N$ by $N$ matrix $U$ as
\beq
 U_a = 
\left(\begin{array}{cccc} 
{\bf 1}_{a-1} & & \\
 &    \cos f_a(r) e^{i \theta} &  - \sin f_a(r) e^{-i\alpha_a} \\
 &    \sin f_a(r)  e^{i\alpha_a} &   \cos f_a(r) e^{-i \theta}\\
 & & & {\bf 1}_{N-a-1} 
\end{array}
\right) 
\quad (a=1,\cdots,N-1). 
\label{eq:skyrme-vortex3}
\eeq
Here, the profile function $f_a$ 
 with the same boundary condition with 
Eq.~(\ref{eq:bc-f}) 
should depends on the masses   $m_a$ and $m_{a+1}$.
The real constant $\alpha_a$ is the $U(1)$ modulus of the $a$-th vortex 
corresponding to that of the $a$-th monopole.

This composite is non-BPS \cite{Eto:2004rz} even in the 
absence of the Josephson term.

\subsection{Monopoles inside a Josephson vortex}

The effect of $\Delta M$ on the effective 
theory of a non-Abelian sine-Gordon soliton, 
that corresponds to 
Josephson vortex, {\it i.e.}, a vortex inside the 
non-Abelian domain wall can be obtained by
the Scherk-Schwarz dimensional reduction 
as before.
The Scherk-Schwarz dimensional reduction
 acts on the moduli $\phi$ 
of the sine-Gordon soliton 
in the exactly same manner with 
those of the non-Abelian vortex in
Eq.~(\ref{eq:SSred-on-vortex}). 
We then obtain the effective theory 
of the non-Abelian sine-Gordon soliton, 
or the Josephson vortex, given by
\beq
{\cal L}_{\rm SG} 
&=& C_X \p_\alpha X \p^\alpha X 
+ C_{\phi}
\left[
\p_\alpha \phi^\dagger \p^\alpha \phi + (\phi^\dagger\p_\alpha \phi)(\phi^\dagger\p^\alpha \phi)
\right] - V \non
 V 
&=&  C_{\phi}
\left[(\phi^\dagger \Delta M \phi)^2-\phi^\dagger (\Delta M)^2 \phi 
\right] .
\eeq
This is again the massive ${\mathbb C}P^{N-1}$ model 
admitting a ${\mathbb C}P^{N-1}$ kink that 
represents a monopole in the bulk.
 We then obtain the configuration in 
Fig.~\ref{fig:monopole-instanton}(e).

A nontrivial consistency check can be done 
from the domain-wall effective theory. 
Let us turn on $\gamma \neq 0$ 
in the domain-wall effective  theory in Eq.~(\ref{eq:eff2}):
\beq
&& {\cal L}_{{\rm wall},\Delta M} =  
{v^2 \over 2m} \del_i X \del^i X 
- {v^2 \over 4m} 
\tr \left(U^\dagger \del_{i} U
            U^\dagger \del^{i} U \right) - V\non
&& 
V = 
   {v^2 \over 4m}  \tr ([ \Delta M, U ]^\dagger[ \Delta M, U ])
 + m'^2  (\tr U + \tr U^\dagger) 
\label{eq:eff3}
\eeq
and consider its effect on the vortex.

\begin{figure}[t]
\begin{center}
\includegraphics[width=0.9\linewidth,keepaspectratio]{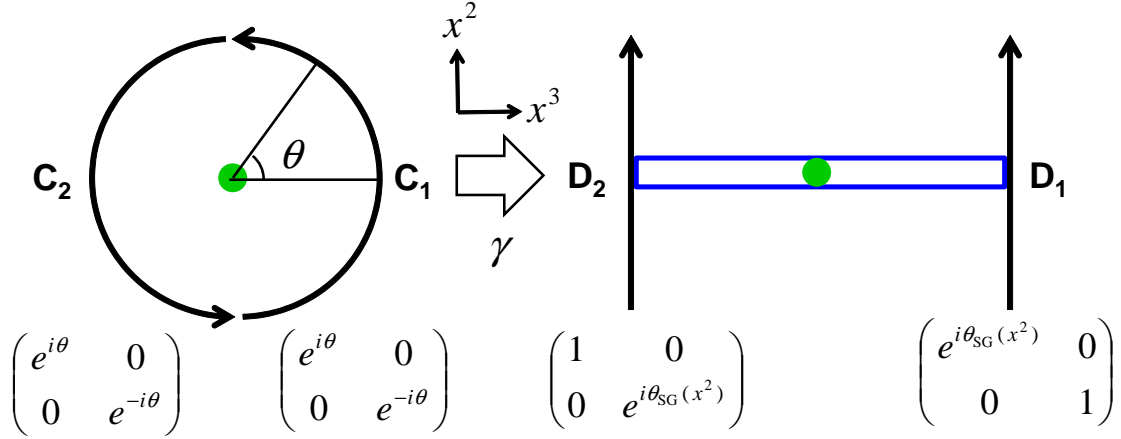}
\caption{
Deformation of a unconfined monopole 
with $\gamma=0$
to a monopole confined by vortices 
with $\gamma\neq 0$, 
corresponding to Fig.~\ref{fig:monopole-instanton} (f) 
and (e), respectively. 
Monopoles are vortices in the $SU(2)$ principal chiral model 
with the mass deformation $\Delta M$ 
realized inside the non-Abelian domain wall 
 (see Fig.~\ref{fig:monopole-instanton}). 
The paths enclosing the vortices are 
the circular path C$_1 +$ C$_2$ for the unconfined monopole (left panel) 
and D$_1 -$ D$_2$ 
(plus the two paths at $x^2 = \pm\infty$ 
where $U$ is constant: $U={\bf 1}_2$) 
for the confined monopole (right panel).
\label{fig:gamma-deformation}}
\end{center}
\end{figure}

For the $N=2$ case, the domain-wall effective theory is
\beq
&& {\cal L}_{{\rm wall},\Delta M} = 
{v^2 \over 4m}\del_i{\bf n}\cdot \del^i{\bf n} 
-  V \non
&& V = {v^2 m_1^2\over 16m}   (n_1^2 + n_2^2) 
+ 2 m'{}^2 n_4.
\eeq
In this case, 
the vortex in Eq.~(\ref{eq:skyrme-vortex})
or (\ref{eq:skyrme-vortex2})
is attached by two sine-Gordon solitons
 with the correct tension $T=8m''$ in Eq.~(\ref{eq:SG-tension}). 
This can be manifest at large distance from the vortex core 
in Eq.~(\ref{eq:skyrme-vortex2}). 
In the absence of $\gamma$, 
the field asymptotically goes to 
\beq
 U \to 
\left(\begin{array}{cc} 
      e^{i \theta} &  0 \\
    0 &  e^{-i \theta} 
\end{array}
\right)    \; {\rm for}\; r\to \infty. 
\label{eq:skyrme-vortex3}
\eeq
In the presence of $\gamma$, this $\theta$ dependence 
should be replaced by two sine-Gordon solitons 
\beq
&& U \to 
\left(\begin{array}{cc} 
    e^{i\theta_{\rm SG}(x^2)} &  0 \\
                                     0 &  1
\end{array}
\right) 
= 
\left(\begin{array}{cc} 
  \exp \left(4 i \, \arctan \exp [m'  (x^2- X)] \right)  &  0 \\
  0 &  1 
\end{array}
\right)  \; {\rm for}\; x^1 \to + \infty, \non
&& U \to
\left(\begin{array}{cc} 
  1 &  0 \\
  0 &  e^{i\theta_{\rm SG}(x^2)}
\end{array}
\right) 
= 
\left(\begin{array}{cc} 
  1 &  0 \\
  0 &  \exp \left(4 i \, \arctan \exp [m'  (x^2- X)] \right) 
\end{array}
\right)  \; {\rm for}\; x^1 \to - \infty, \non
&& U \to {\bf 1}_2 \; {\rm for}\; x^2 \to \pm \infty,
\label{eq:skyrme-vortex4}
\eeq
where we have used Eq.~(\ref{eq:U(1)-one-kink}).
This deformation is illustrated in 
Fig.~\ref{fig:gamma-deformation}.
Note that 
the direction of the path D$_2$ at $x^1 \to -\infty$
in Eq.~(\ref{eq:skyrme-vortex4}) 
is opposite to the left side  C$_2$ of the angular path 
C$_1$+C$_2$ in Eq.~(\ref{eq:skyrme-vortex3}) 
so that the lower-right components of $U$'s in  
Eqs.~(\ref{eq:skyrme-vortex3}) and 
(\ref{eq:skyrme-vortex4}) 
have the same windings.
These sine-Gordon solitons carry opposite 
${\mathbb C}P^1$ moduli.
The sine-Gordon solitons are vortices 
from the bulk point of view. 
Therefore, we have shown that 
the monopole must be confined by 
the two vortices with the opposite 
${\mathbb C}P^1$ moduli.
 We then again reach the configuration in 
Fig.~\ref{fig:monopole-instanton}(e).

\begin{figure}[t]
\begin{center}
\includegraphics[width=0.9\linewidth,keepaspectratio]{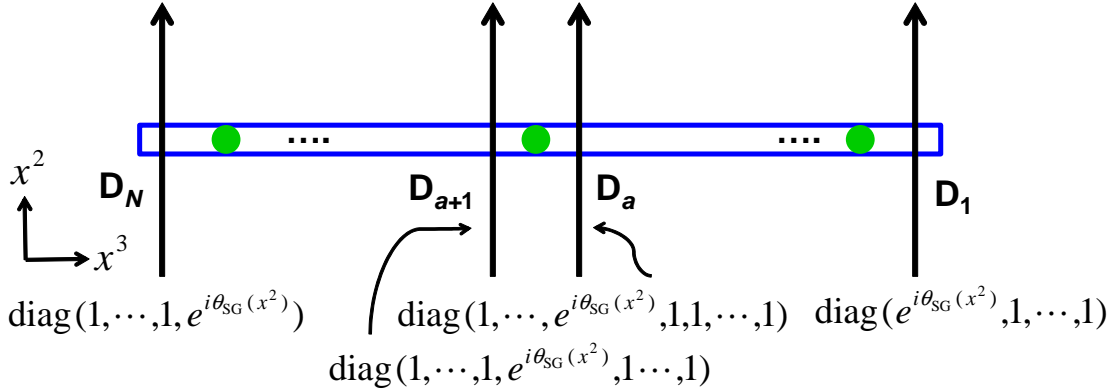}
\caption{
Confined monopoles are vortices in the $SU(N)$ principal chiral model 
with the mass deformation $\Delta M$ 
realized inside the non-Abelian domain wall 
 (see Fig.~\ref{fig:monopole-instanton}). 
The path enclosing the $a$-th vortex is 
D$_a -$ D$_{a+1}$ 
(plus the two paths at $x^2 = \pm\infty$ 
where $U$ is constant: $U={\bf 1}_N$). 
\label{fig:gamma-deformation2}}
\end{center}
\end{figure}

For general $N$, the asymptotic form 
of the $a$-th vortex becomes 
\beq
 U_a \to 
\left(\begin{array}{cccc} 
{\bf 1}_{a-1} & & \\
 &    e^{i \theta} &  0 \\
 &    0  &   e^{-i \theta}\\
 & & & {\bf 1}_{N-a-1} 
\end{array}
\right) 
\quad (a=1,\cdots,N-1)
\quad {\rm for}\; r \to \infty 
\label{eq:skyrme-vortex5}
\eeq
 for $\gamma=0$.
In the presence of $\gamma$, this $\theta$ dependence 
should be replaced by two sine-Gordon solitons as before:
\beq
&& 
 U_a \to
\left(\begin{array}{cccc} 
{\bf 1}_{a-1} & & \\
 &    e^{i\theta_{\rm SG}(x^2)}  &  0 \\
 &    0  &   1 \\
 & & & {\bf 1}_{N-a-1} 
\end{array}
\right)  \; {\rm for}\; x^1 \to + \infty, \non
&& U_a \to
\left(\begin{array}{cccc} 
{\bf 1}_{a-1} & & \\
 &    1 &  0 \\
 &    0  &   e^{i\theta_{\rm SG}(x^2)}  \\
 & & & {\bf 1}_{N-a-1} 
\end{array}
\right)  
 \; {\rm for}\; x^1 \to - \infty, \non
&& U \to {\bf 1}_N \; {\rm for}\; x^2 \to \pm \infty.
\label{eq:skyrme-vortex6}
\eeq
For composite solitons, 
the first and second paths are 
D$_a$ and D$_{a+1}$, respectively, in 
Fig.~\ref{fig:gamma-deformation2}.

The wall-vortex-monopole composites studied here are non-BPS. 
In general, wall-vortex-monopole composites can be 1/4 BPS 
if embedded into a supersymmetric theory,
only when the vortices are perpendicular to the domain wall 
\cite{Isozumi:2004vg,Eto:2005sw}.

As for instantons, 
we have two limits to remove host solitons.\\
(1) $m\to 0$: 
the non-Abelian domain wall disappears in this limit,  
since the width of it is $m^{-1}$.
The configuration is a vortex-monopole composite,
where monopoles are ${\mathbb C}P^{N-1}$ 
kinks in the vortex. 
This composite is 1/4 BPS if embedded into a supersymmetric theory 
\cite{Eto:2004rz,Eto:2005sw}. 
\\
(2) $\gamma \to 0$: 
the non-Abelian vortex disappears in this limit, 
since the size of the vortex along the domain wall 
world-volume is proportional to $m''^{-1} \sim v/\sqrt{\gamma}$.
The configuration is a wall-monopole composite,
where monopoles are vortices in the wall.
This composite is non-BPS \cite{Eto:2005sw}.
Therefore, the original configuration 
gives a duality between 
${\mathbb C}P^{N-1}$ kinks
and $U(1)^{N-1}$ vortices
both realized by Yang-Mills instantons.

\section{Summary and Discussion \label{sec:summary} }

We have studied instantons and monopoles in
a non-Abelian Josephson junction, 
that is, a junction of 
non-Abelian color superconductors 
sandwiching an insulator. 
Low-energy dynamics of 
a non-Abelian domain wall   
can be described by
a $U(N)$ principal chiral model, 
where a non-Abelian Josephson vortex,
a non-Abelian vortex (color magnetic flux tube) 
residing inside the junction,  
is described as a non-Abelian sine-Gordon soliton.  
Josephson instantons and monopoles 
have been realized inside the non-Abelian 
Josephson vortex inside the junction.
By removing the junction with the vanishing Higgs mass, 
$m=0$,  
the configurations go back to the well-known 
instanton-vortex and monopole-vortex composites. 
On the other hand, if we remove the vortex by 
turning off the Josephson coupling $\gamma$,  
there remain (unconfined) instantons and monopoles 
inside the junction (instanton-wall and monopole-wall composites). 
The whole situation is illustrated in 
Fig.~\ref{fig:monopole-instanton}.
We have found 
that monopoles become $U(1)^{N-1}$ vortices 
in the $U(N)$ principal chiral model  
inside the junction, with the matching of 
the monopole charge 
$\pi_2[SU(N)/U(1)^{N-1}] \simeq {\mathbb Z}^{N-1}$ 
and the vortex charge 
$\pi_1[U(1)^{N-1}] \simeq {\mathbb Z}^{N-1}$,
while it was known that instantons become Skyrmions there.
We have confirmed the monopole confinement 
in the junction; when we turn on the Josephson 
coupling $\gamma$, the $U(1)^{N-1}$ vortices 
must be confined by the non-Abelian 
sine-Gordon solitons, implying that 
the monopole must be confined by the non-Abelian vortices 
in the bulk point of view.
As summarized in Table~\ref{table:SS-dim-red},
we have shown that 
the T-duality relations between instantons and monopoles 
induces the duality 
between the $SU(N)$ Skyrmions and 
the $U(1)^{N-1}$ vortices (inside the junction)
as well as the previously-known duality between 
 ${\mathbb C}P^{N-1}$ lumps 
and ${\mathbb C}P^{N-1}$ kinks (inside the vortex). 
We have also observed 
the new kind of duality between  ${\mathbb C}P^{N-1}$ lumps 
and $SU(N)$ Skyrmions 
and that between ${\mathbb C}P^{N-1}$ kinks and 
$U(1)^{N-1}$ vortices, 
as well as that between fractional instantons and bions 
in the ${\mathbb C}P^{N-1}$ model in 
two Euclidean dimensions 
and those in the $SU(N)$ principal chiral model in 
three Euclidean dimensions.

When we add the Chern-Simmons term in the gauge theory in the 4+1 dimensional bulk, the Wess-Zumino-Witten term is  induced on the domain wall world-volume.
This term would be important to interpret 
the domain wall world-volume theory 
as the low-energy effective theory of QCD.

In this paper, we have  considered $U(N)$ gauge theory 
but $SU(N)$ gauge group does not change the main results,
implying that those 
can be applied to color superconductors 
appearing in high density quark matter 
\cite{Alford:2007xm,Eto:2013hoa},
where non-Abelian vortices are superfluid vortices 
with color magnetic fluxes confined inside their cores 
\cite{Balachandran:2005ev}. 
If quark matter is separated by an insulator 
for instance by some modulation such as crystalline superconductivity, 
it will give (an array of) non-Abelian Josephson junctions.
Non-Abelian vortices, 
monopoles and instantons there 
become non-Abelian Josephson vortices, 
Josephson monopoles and Josephson instantons, 
respectively,  
by trapped inside the insulating region. 

As mentioned in introduction 
our configurations suggest a duality 
between fractional instantons and bions 
in the ${\mathbb C}P^{N-1}$ model on ${\mathbb R}^1\times S^1$ 
and the $SU(N)$ principal chiral model 
on ${\mathbb R}^2\times S^1$ 
with twisted boundary conditions. 
Hopefully this duality may be useful to understand 
the resurgence of these models 
together with $SU(N)$ Yang-Mills theory on ${\mathbb R}^3\times S^1$ 
at quantum level 
from a unified point of view.

\begin{figure}[t]
\begin{center}
\includegraphics[width=0.9\linewidth,keepaspectratio]{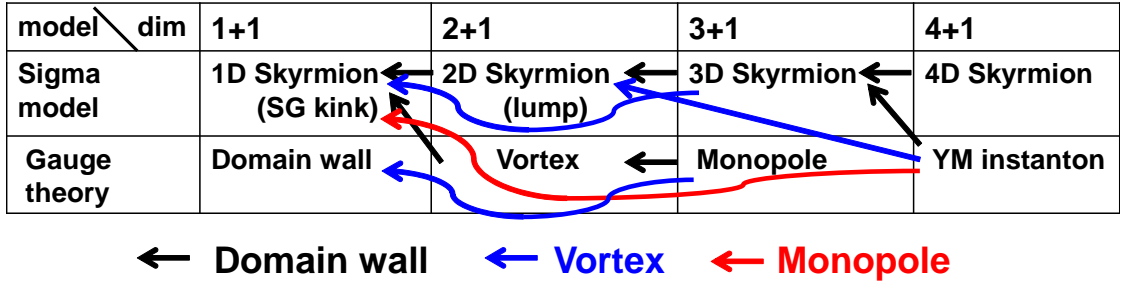}
\caption{
Unification of topological solitons and instantons.
The starting points of arrows are solitons in the bulk 
which become the endpoints of arrows when 
reside inside host solitons.
The length of the arrows denote the host solitons: 
The black, blue and red arrows 
connecting one, two and three columns 
denote a domain wall, vortex and monopole. 
\label{fig:unification}}
\end{center}
\end{figure}

Let us mention what the results in this paper 
imply for unified understanding of topological 
solitons and instantons.
Various relations between host and 
daughter solitons found thus far 
are summarized in Fig.~\ref{fig:unification}.
Our new finding here is that a monopole 
becomes a vortex inside 
a non-Abelian domain wall.

\section*{Acknowledgements}

The author thanks Dereck Harland for bringing his attention to 
Ref.~\cite{Harland:2008eu} 
and Minoru Eto and Sven Bjarke Gudnason for collaborations 
in related works. 
This work is supported in part by 
Grant-in-Aid for Scientific Research (No. 23740198) 
and by the ``Topological Quantum Phenomena'' 
Grant-in-Aid for Scientific Research 
on Innovative Areas (No. 23103515)  
from the Ministry of Education, Culture, Sports, Science and Technology 
(MEXT) of Japan.


\newcommand{\J}[4]{{\sl #1} {\bf #2} (#3) #4}
\newcommand{\andJ}[3]{{\bf #1} (#2) #3}
\newcommand{\AP}{Ann.\ Phys.\ (N.Y.)}
\newcommand{\MPL}{Mod.\ Phys.\ Lett.}
\newcommand{\NP}{Nucl.\ Phys.}
\newcommand{\PL}{Phys.\ Lett.}
\newcommand{\PR}{ Phys.\ Rev.}
\newcommand{\PRL}{Phys.\ Rev.\ Lett.}
\newcommand{\PTP}{Prog.\ Theor.\ Phys.}
\newcommand{\hep}[1]{{\tt hep-th/{#1}}}


\begin{thebibliography}{100}
\bibitem{footnote2}
In a supersymmetric extension of our model, 
the full moduli space is larger than that in Eq.~(\ref{eq:wall-moduli}); 
It is $T^* U(N)$, that is 
the cotangent bundle over $U(N)$ and is K\"ahler.
\end{thebibliography}
\end{document}